\documentclass[%
 aip,
 jmp,%
 amsmath,amssymb,
 reprint,%
]{revtex4-1}

\usepackage{graphicx}
\usepackage{dcolumn}
\usepackage{bm}

\begin{document}


\title[Transient solution for vesicle electrodeformation and relaxation]{A transient solution for vesicle electrodeformation and relaxation}

\author{Jia Zhang,$^1$ Jeffery D. Zahn,$^2$ Wenchang Tan,$^3$ and Hao Lin$^{}$}
\thanks{Email address for correspondence: hlin@jove.rutgers.edu}
\affiliation{$^1$Department of Mechanical and Aerospace Engineering, Rutgers, The State University of New Jersey, Piscataway, NJ 08854, USA \\
$^2$Department of Biomedical Engineering, Rutgers, The State University of New Jersey, Piscataway, NJ 08854, USA\\
$^3$State Key Laboratory for Turbulence and Complex Systems, Department of Mechanics and Engineering Science, College of Engineering, Peking University, Beijing 100871, China}%

\date{\today}

\begin{abstract}
A transient analysis for vesicle deformation under DC electric fields
is developed. The theory extends from a droplet model, with the additional
consideration of a lipid membrane separating two fluids of arbitrary
properties. For the latter, both a membrane-charging and a membrane-mechanical
model are supplied. The vesicle is assumed to remain spheroidal in
shape for all times. The main result is an ODE governing the evolution
of the vesicle aspect ratio. The effects of initial membrane tension
and pulse length are examined. The model prediction is extensively
compared with experimental data, and is shown to accurately capture
the system behavior in the regime of no or weak electroporation. More
importantly, the comparison reveals that vesicle relaxation obeys
a universal behavior regardless of the means of deformation. The process
is governed by a single timescale that is a function of the vesicle initial
radius, the fluid viscosity, and the initial membrane tension. This
universal scaling law can be used to calculate membrane properties
from experimental data. 
\end{abstract}

\maketitle

\section{Introduction}

Vesicles are widely used as a model system for biological cells due
to their simplicity and controllability. The deformation of the lipid
membrane, in particular under an applied electric field (electrodeformation),
is often explored to probe membrane properties \cite{Kummrow1991,Niggemann1995} and to detect pathological changes in cells.
\cite{Wong2005}

In the past decade, vesicle electrodeformation has become a significant
subject of study, and earlier work can be divided into two categories.
In the first category, an alternating-current (AC) field is applied, which often induces
stationary and small deformations. \cite{Kummrow1991,Niggemann1995,Dimova2007,Aranda2008} Correspondingly, an electrohydrodynamic
theory in the small-deformation limit was developed to interpret
the data trends. \cite{Vlahovska2009} In the second, under direct-current (DC)
electric fields, vesicles usually exhibit large and transient deformations
due to the large field strengths commonly applied. \cite{Kakorin2003,Riske2005,Riske2006,Sadik2011} Recently,
using high-resolution, high-speed optical imaging Riske and Dimova\cite{Riske2005}
acquired a large amount of data capturing the complex deformation-relaxation
behavior of the vesicles. Although some qualitative and scaling arguments
were presented, \cite{Dimova2007} the data was not fully interpreted
due to the absence of a predictive model. Meanwhile, one of us (HL)
experimentally examined vesicles in the large-deformation regime with
aspect ratios reaching ten. \cite{Sadik2011} A large-deformation
theory was also presented, which provided quantitative agreement with
the data therein. However, the model was semi-empirical in that the
hydrodynamic problem was not rigorously treated, but followed an empirical
approach by Hyuga and co-authors. \cite{Hyuga1991a,Hyuga1991}
In general, a rigorous and transient analysis needs to be developed
to understand the complex deformation-relaxation behavior, and to
provide insights on the underlying physical processes.

In this work, we develop a transient analysis for vesicle electrodeformation.
The theory is derived by extending our previous work on a droplet model, \cite{Zhang2012} with the additional
consideration of a lipid membrane separating two fluids of arbitrary
properties. For the latter, both a membrane-charging and a membrane-mechanical
model are supplied. Similar to the droplet model, the main result
is also an ordinary differential equation (ODE) governing the evolution of the vesicle aspect ratio.
The effects of initial membrane tension and pulse length are examined.
The model prediction is extensively compared with experimental data
from Riske and Dimova \cite{Riske2005} and Sadik \emph{et al.},\cite{Sadik2011} and is shown to accurately
capture the system behavior in the regime of no or weak electroporation.
More importantly, the comparison reveals that vesicle relaxation obeys
a universal behavior, and is governed by a single timescale that is
a function of the vesicle initial radius, the fluid viscosity, and the
initial membrane tension. This behavior is regardless of the means
of deformation, either via AC/DC electric field, or via mechanical
stretching. This universal scaling law is a main contribution of the
current work, and can be used to calculate membrane properties from
experimental data. 

\section{Theory}

\begin{figure}
\center\includegraphics[width=0.35\textwidth]{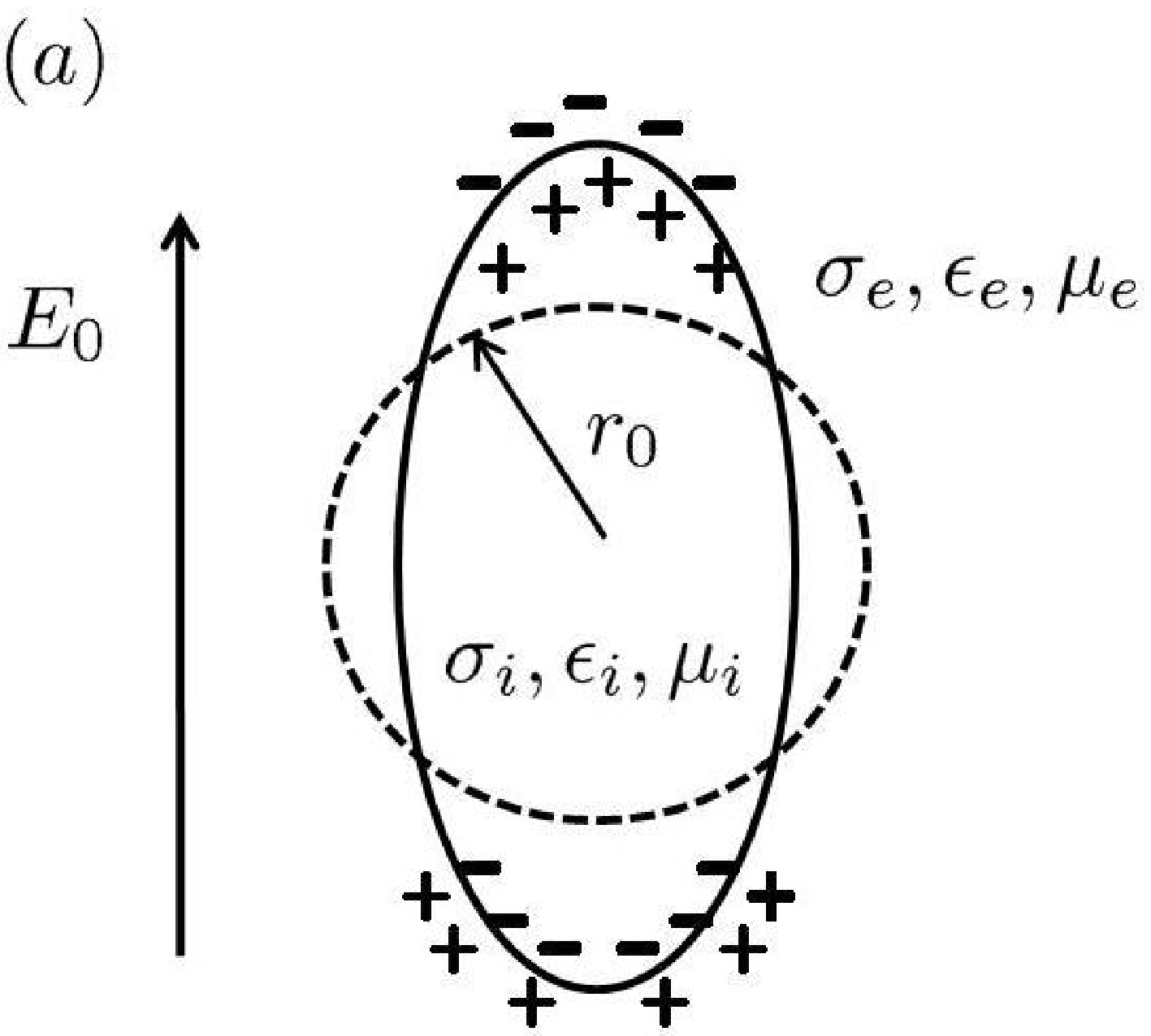}
\includegraphics[width=0.35\textwidth]{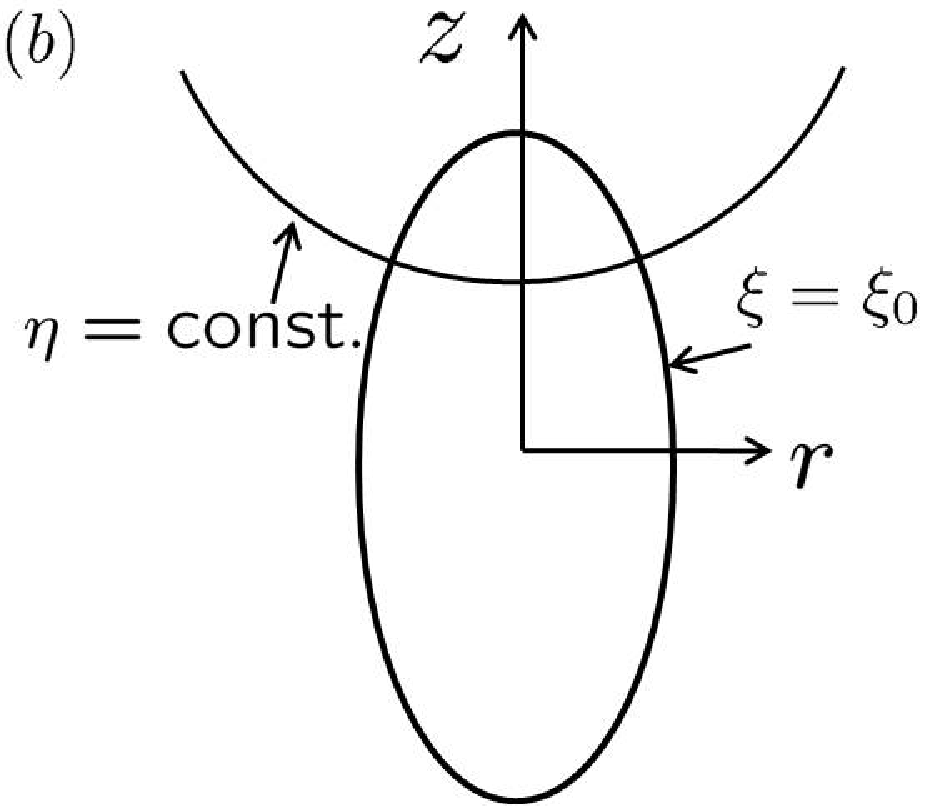}
\caption{(a) A schematic of the problem configuration. The original radius of the
vesicle is $r_{0}$. The conductivity is denoted by $\sigma$, the
permittivity is denoted by $\epsilon$, the viscosity is denoted by
$\mu$, and the subscripts $i$ and $e$ denote intravesicular and
extravesicular, respectively. The strength of the applied electric
field is $E_{0}$. (b) The prolate spheroidal
coordinate system. \label{fig:Schematics-of-problem.}}

\end{figure}

The problem configuration is shown in Fig. \ref{fig:Schematics-of-problem.}.  Under the influence of an applied electric field,
charges of opposite signs are allowed to accumulate on the two sides
of the membrane, which induces vesicle deformation and electrohydrodynamic flows both inside and outside the vesicle. We assume that the vesicle remains spheroidal in shape throughout the process. All notations, as well as the prolate spheroidal
coordinate system follow those from Zhang \emph{et al.}.\cite{Zhang2012} The surface of the prolate spheriod is conveniently given as\begin{equation}
\xi=\xi_{0}\equiv\frac{a}{c}.\label{prolate surface}\end{equation}Here $c=\sqrt{a^{2}-b^{2}}$ is chosen to be the semi-focal length
of the spheroidal vesicle, and $a$ and $b$ are the major and minor
semi-axis, respectively. For the derivation below, we further assume that the volume of the
vesicle is conserved. We subsequently obtain\begin{equation}
a=r_{0}(1-\xi_{0}^{-2})^{-\frac{1}{3}},\qquad b=r_{0}(1-\xi_{0}^{-2})^{\frac{1}{6}}.\label{ab}\end{equation}
Therefore, the vesicle geometry is completely characterized by a single
parameter, $\xi_{0}$, which evolves in time along with deformation.
The critical idea of the current analysis is to express all variables,
e.g., the electric potential and the stream function in terms of $\xi_{0}$.
In what follows, we introduce both an electrical
and a mechanical model for the membrane. An ODE for $\xi_{0}$ is obtained by applying the stress matching and kinematic conditions.

\subsection{The electrical problem}

The electric potentials both inside and outside the vesicle are described by the Laplace equations:\begin{equation}
\nabla^{2}\phi_{i}=\nabla^{2}\phi_{e}=0.\label{laplace for vesicle}\end{equation}
However, at the membrane the matching conditions are modified:\begin{eqnarray}
\frac{\sigma_{e}}{h_{\xi}}\frac{\partial\phi_{e}}{\partial\xi}=\frac{\sigma_{i}}{h_{\xi}}\frac{\partial\phi_{i}}{\partial\xi}=&&C_{m}\frac{\partial\frac{c}{h_{\xi}}(\phi_{e}-\phi_{i})}{\partial t}\nonumber\\
&&+\frac{G_{m}c}{h_{\xi}}(\phi_{e}-\phi_{i}),\qquad{\rm at}\ \xi=\xi_{0}.\label{eq:current continous}\end{eqnarray}
Here $C_{m}$ and $G_{m}$ denote the membrane capacitance and conductance,
respectively. $h_{\xi}$ is a metric coefficient of the prolate spheroidal
coordinate system. This membrane-charging model is commonly adopted by
many previous research. \cite{Schwan1989,Grosse1992,Debruin1999,Krassowska2007,Li2011} The displacement currents from the electrolytes
are not included, which approximation is valid when the Maxwell-Wagner
timescale, $T_{MW}=(\epsilon_{i}+2\epsilon_{e})/(\sigma_{i}+2\sigma_{e})$,
and the charge relaxation timescale, $T_{cr}=\epsilon/\sigma$, are
small when compared with the membrane-charging time, $T_{ch}=r_{0}C_{m}(1/\sigma_{i}+1/2\sigma_{e})$,
and the deformation time, $T_{d}=\mu_{e}/\epsilon_{e}E_{0}^{2}$.
However, the last two times are in general comparable with each other.
The first term on the RHS of Eq. (\ref{eq:current continous})
represents capacitive charging of the membrane, which includes the
effect of membrane deformation. However, the contribution from this
effect is usually small, and is neglected in the current analysis
for simplicity. Equation (\ref{eq:current continous}) can be consequently
reduced to\begin{eqnarray}
\frac{\sigma_{e}}{h_{\xi}}\frac{\partial\phi_{e}}{\partial\xi}=\frac{\sigma_{i}}{h_{\xi}}\frac{\partial\phi_{i}}{\partial\xi}=&&\frac{C_{m}c}{h_{\xi}}\frac{\partial(\phi_{e}-\phi_{i})}{\partial t}\nonumber\\
&&+\frac{G_{m}c}{h_{\xi}}(\phi_{e}-\phi_{i}),\qquad{\rm at}\ \xi=\xi_{0}.\label{eq:reduced current continous}\end{eqnarray}
Equation (\ref{eq:reduced current continous}) can be further simplified
by considering different stages of charging. In the first stage, the
transmembrane potential (TMP), $V_{m}\equiv(\phi_{i}-\phi_{e})_{\xi=\xi_{0}}$,
grows continuously in magnitude, but the membrane is not permeabilized.
Under this condition, $G_{m}$ is near zero, and Eq. (\ref{eq:reduced current continous})
becomes\begin{equation}
\frac{\sigma_{e}}{h_{\xi}}\frac{\partial\phi_{e}}{\partial\xi}=\frac{\sigma_{i}}{h_{\xi}}\frac{\partial\phi_{i}}{\partial\xi}=\frac{C_{m}c}{h_{\xi}}\frac{\partial(\phi_{e}-\phi_{i})}{\partial t},\qquad{\rm at}\ \xi=\xi_{0}.\label{eq:capacitive current continous}\end{equation}
In the second stage, the maximum TMP reaches the critical threshold,
$V_{c}$, for electroporation to occur.\cite{Chang1990,Leontiadou2004,Gurtovenko2005,Tarek2005,Wohlert2006,Pliquett2007,Fernandez2010} The membrane becomes
permeable to ions, and $G_{m}$ increases significantly to limit further
growth of the TMP. In general, the exact values of $V_{m}$ and $G_{m}$
depend on the detailed electroporation conditions and variables such
as pore density and pore area. \cite{Li2011} The solution usually
requires a complex numerical simulation which is beyond the scope
of the theoretical analysis pursued in this paper. However, a comprehensive
model study by Li and Lin \cite{Li2011} showed that the maximum TMP remained
at the critical level in the presence of the pulse post-permeabilization.
In this work, we adopt an approximate model for this stage. We assume
that once the maximum value of $V_{m}$ reaches $V_{c}$, it no longer
grows and {}``freezes'' in time. In addition, the membrane is completely
permeabilized, and Eq. (\ref{eq:reduced current continous})
is replaced by\begin{equation}
\frac{\sigma_{e}}{h_{\xi}}\frac{\partial\phi_{e}}{\partial\xi}=\frac{\sigma_{i}}{h_{\xi}}\frac{\partial\phi_{i}}{\partial\xi},\quad V_{m}=V_{c},\qquad{\rm at}\ \xi=\xi_{0}.\label{eq:reach critical Vm}\end{equation}
Note that electroporation only occurs for sufficiently strong electric
fields, and Eq. (\ref{eq:reach critical Vm}) is not needed for
some of the cases studied below where $V_{c}$ is never reached. Far
away from the vesicle surface, the electric field is uniform\begin{equation}
-\nabla\phi_{e}=E_{0}\mathbf{\boldsymbol{z}},\qquad{\rm at}\ \xi\rightarrow\infty.\label{eq:farfield electric vesicle}\end{equation}
We also require that $\phi_{i}$ remains finite at $\xi=1$. For initial
condition, we solve Eqs. (\ref{laplace for vesicle}) and (\ref{eq:capacitive current continous})
with $V_{m}=0$.

The general solution of the electric potentials for both the exterior
and interior of the vesicle can be obtained following a similar procedure
outlined in Zhang \emph{et al.}:\cite{Zhang2012} \begin{equation}
\phi_{e}=E_{0}r_{0}\left[-\lambda\xi+\alpha Q_{1}(\xi)\right]\eta,\label{exterior potential}\end{equation}
\begin{equation}
\phi_{i}=E_{0}r_{0}\beta\xi\eta.\label{interior potential}\end{equation}
Here, $Q_{1}(\xi)$ is a 1st-degree Legendre polynomial of the second
kind. $\lambda\equiv c/r_{0}$ is the dimensionless semi-focal length.
The coefficients $\alpha$ and $\beta$ are again obtained by applying
the matching conditions. In the absence of electroporation, they are
given as \begin{equation}
\alpha=\frac{\beta+\sigma_{r}\lambda}{Q_{1}^{'}(\xi_{0})\sigma_{r}},\label{a(t)}\end{equation}
\begin{widetext}
\begin{eqnarray}
\left[\frac{Q_{1}(\xi_{0})}{Q_{1}^{'}(\xi_{0})\sigma_{r}}-\xi_{0}\right]\frac{d\beta}{d\tau}-&&\left[\frac{Q_{1}(\xi_{0})Q_{1}^{''}(\xi_{0})-Q_{1}^{'2}(\xi_{0})(1-\sigma_{r})}{Q_{1}^{'2}(\xi_{0})\sigma_{r}}\frac{d\xi_{0}}{d\tau}+\frac{\tau_{2}}{\tau_{1}\lambda}\right]\beta\nonumber\\
&&-\left[\left(\xi_{0}-\frac{Q_{1}(\xi_{0})}{Q_{1}^{'}(\xi_{0})}\right)\frac{d\lambda}{d\xi_{0}}+\frac{\lambda Q_{1}^{''}(\xi_{0})Q_{1}(\xi_{0})}{Q_{1}^{'2}(\xi_{0})}\right]\frac{d\xi_{0}}{d\tau}=0,\label{b(t)}
\end{eqnarray}
\begin{equation}
\alpha(0)=\frac{\lambda\xi_{0}(\sigma_{r}-1)}{Q_{1}^{'}(\xi_{0})\xi_{0}\sigma_{r}-Q_{1}(\xi_{0})},\qquad\beta(0)=\left[-\lambda+\alpha(0)Q_{1}^{'}(\xi_{0})\right]\sigma_{r}.
\label{initial ab}\end{equation}
\end{widetext}
Here $\sigma_{r}\equiv\sigma_{e}/\sigma_{i}$ is the conductivity
ratio. $\tau_{1}\equiv r_{0}C_{m}/\sigma_{i}$ is a membrane-charging
time. $\tau_{2}\equiv r_{0}\mu_{e}/\Gamma_{0}$ is a characteristic
flow timescale. $\Gamma_{0}$ is the initial membrane tension introduced
below. The dimensionless time $\tau$ defined as $\tau\equiv t/\tau_{2}$
has been used. Note that the definition of these times slightly deviates
from those used in Zhang \emph{et al.}\cite{Zhang2012} due to the difference between
droplet and vesicle. However, $\tau_{2}$ remains formally the same
by replacing $\gamma$ in Zhang \emph{et al.}\cite{Zhang2012} with $\Gamma_{0}$.

After the maximum value of $V_{m}$ reaches the critical threshold,
electroporation occurs. $\alpha$ and $\beta$ are calculated by Eq.
(\ref{eq:reach critical Vm}) which yields \begin{equation}
\alpha=\frac{-V_{c}/(E_{0}r_{0})-\lambda\xi_{0}(\sigma_{r}-1)}{Q_{1}(\xi_{0})-Q_{1}^{'}(\xi_{0})\xi_{0}\sigma_{r}},\:\beta=\left[-\lambda+\alpha Q_{1}^{'}(\xi_{0})\right]\sigma_{r}.\label{electroporated ab}\end{equation}

The expressions for the normal and tangential electrostatic stresses
are found in Zhang \emph{et al.} \cite{Zhang2012} and not repeated here.

\subsection{The hydrodynamic problem}

In the regime of low-Reynolds-number flow, the governing equation
for the hydrodynamic problem can be rewritten in terms of the stream
function, $\psi$, as \begin{equation}
\rm{E}^{4}\psi=0.\label{stream function}\end{equation}
Here, the expression for the operator $\rm{E}^{2}$ can be found in Dubash and Mestel\cite{Dubash2007} and Bentenitis and Krause.\cite{Bentenitis2005} The stream function is related to the velocity components as\begin{equation}
u=-\frac{1}{h_{\xi}h_{\theta}}\frac{\partial\psi}{\partial\xi},\qquad v=\frac{1}{h_{\eta}h_{\theta}}\frac{\partial\psi}{\partial\eta}.\label{velocity field}\end{equation}
$h_{\eta}$ and $h_{\theta}$ are metric coefficients of the prolate spheroidal coordinate
system. At the membrane, $u$ and $v$ represent the tangential and
normal velocities, respectively, and they are required to be continuous
\begin{equation}
u_{e}=u_{i},\qquad v_{e}=v_{i},\qquad{\rm at}\ \xi=\xi_{0}.\label{eq:no-slip condition}\end{equation}
In addition, we prescribe a kinematic condition relating the membrane
displacement to the normal velocity, \begin{equation}
v(\xi=\xi_{0},\:\eta)=\frac{r_{0}\left(1-\xi_{0}^{-2}\right)^{-5/6}}{3\xi_{0}^{2}}\frac{\left(1-3\eta^{2}\right)}{\sqrt{\xi_{0}^{2}-\eta^{2}}}\frac{d\xi_{0}}{dt}.\label{eq:kinematic equation}\end{equation} At the membrane, the stress matching condition is given as:\begin{equation}
||\tau\cdot\mathbf{\boldsymbol{n}}||=\mathbf{\boldsymbol{f}}^{mem}.\label{eq:stress balance}\end{equation}
Here $\mathbf{\boldsymbol{f}}^{mem}$ is the surface force density
arising from the vesicle membrane. The tensor $\tau$ includes contributions
from both the hydrodynamic and electrostatic stresses:\begin{equation}
\tau\equiv-p{\rm I}+\mu(\nabla\mathbf{\boldsymbol{v}}+\nabla\mathbf{\boldsymbol{v}}^{T})+\epsilon\mathbf{\boldsymbol{E}}\mathbf{\boldsymbol{E}}-\frac{1}{2}\epsilon(\mathbf{\boldsymbol{E}}\cdot\mathbf{\boldsymbol{E}}){\rm I}.\label{eq:stress}\end{equation}

\begin{figure}
\center

\includegraphics[width=0.5\textwidth]{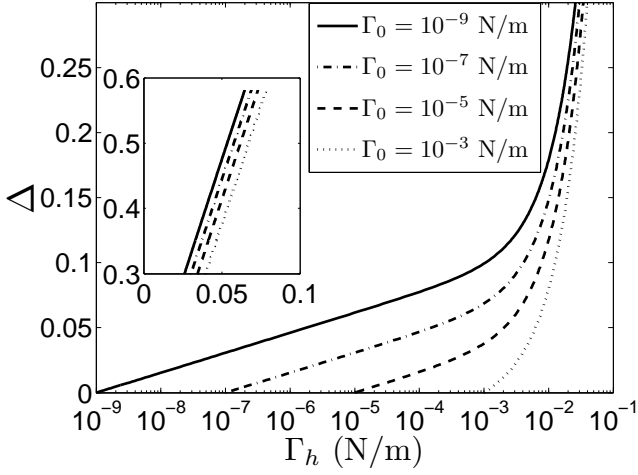}

\caption{The relative increase of the apparent area, $\Delta$, as a function
of membrane tension, $\Gamma_{h}$, for different values of initial
membrane tension, $\Gamma_{0}$. The inset shows the linear regime
for larger $\Gamma_{h}$ values.\label{fig:Relative-area-expansion}}

\end{figure}

\subsection{The membrane-mechanical model}

The surface force density at the vesicle membrane essentially consists
of two parts \cite{Seifert1997,Vlahovska2009}\begin{equation}
\mathbf{\boldsymbol{f}}^{mem}=\mathbf{\boldsymbol{f}}^{\kappa}+\mathbf{\boldsymbol{f}}^{\Gamma}.\label{eq:surface force density}\end{equation}
Here $\mathbf{\boldsymbol{f}}^{\kappa}$ is the surface force density
induced by bending resistance. $\mathbf{\boldsymbol{f}}^{\Gamma}=2\Gamma H\mathbf{\boldsymbol{n}}-\nabla_{s}\Gamma$
is the surface force density induced by the membrane tension. $H$
is the mean curvature, and $\Gamma$ is the local membrane tension.
We can easily verify that $\mathbf{\boldsymbol{f}}^{\kappa}$ is several
orders of magnitude smaller than $\mathbf{\boldsymbol{f}}^{\Gamma}$,
and is therefore not included in the current analysis. The local membrane
tension, $\Gamma$, is calculated by assuming an effective tension
which is uniform over the entire membrane.\cite{Helfrich1984,Vlahovska2009} An increase of the homogeneous tension, $\Gamma_{h}$,
from the initial tension, $\Gamma_{0}$, leads to an increase in the
apparent membrane area: \cite{Helfrich1984,Evans1990,Kummrow1991,Evans1991}\begin{equation}
\Delta=\frac{k_{B}T}{8\pi\kappa}{\rm ln}\frac{\Gamma_{h}}{\Gamma_{0}}+\frac{\Gamma_{h}-\Gamma_{0}}{K_{a}}.\label{membrane tension}\end{equation}
Here $\Delta$ is the increase in the apparent membrane area relative
to the initial spherical state,\begin{equation}
\Delta=\frac{1}{2}\left(1-\xi_{0}^{-2}\right)^{-\frac{2}{3}}\left[1-\xi_{0}^{-2}+\left(\xi_{0}^{2}-1\right)^{\frac{1}{2}}{\rm arcsin}\left(\xi_{0}^{-1}\right)\right]-1.\label{eq:relative area increase}\end{equation}
$K_{a}$ is the elastic stretching modulus. $\kappa$ is the bending
rigidity. Equation (\ref{membrane tension}) indicates that $\Gamma_{0},\:\kappa$,
and $K_{a}$ are the important parameters in determining membrane
tension. $\kappa$ and $K_{a}$ are usually constants for a specific
vesicle type, and their values are often readily obtained from previous
work. \cite{Kwok1981,Kummrow1991,Needham1995}
On the other hand, $\Gamma_{0}$ is specific to an individual vesicle,
and its value can not be directly determined from experimental measurements.
The relation between $\Delta$ and $\Gamma_{h}$ for different choices
of $\Gamma_{0}$ is shown in Fig. \ref{fig:Relative-area-expansion}.
When $\Delta$ is small, the membrane area increases through the flattening
of the undulations, and $\Gamma_{h}$ shows an exponential correlation
with $\Delta$. When $\Delta$ is sufficiently large, a linear behavior
is observed instead, and the membrane area increase is mainly due
to elastic stretching. Moreover, a larger $\Gamma_{0}$ always leads
to a larger $\Gamma_{h}$ for the same value of $\Delta$. 

\subsection{General solution}

A solution for vesicle electrodeformation can be obtained by solving
the governing equations of both the electrical and hydrodynamic problems,
with the help of the matching conditions. The solution strategy is identical
to that presented in Zhang \emph{et al.},\cite{Zhang2012} with only differences in
the detailed matching conditions for both the electric field and the
interfacial forces. For brevity, only the final governing equation
for $\xi_{0}$ is presented here:
\begin{widetext}
\addtocounter{equation}{0}\begin{subequations}\begin{equation}
\frac{d\xi_{0}}{d\tau}=-\frac{1}{F}\left[Q_{N}f_{21}(\xi_{0})+Q_{T}\frac{\mu_{r}f_{22}(\xi_{0})+f_{23}(\xi_{0})}{\mu_{r}f_{14}(\xi_{0})+f_{15}(\xi_{0})}-\frac{\Gamma_{h}}{\Gamma_{0}}f_{24}(\xi_{0})\right],\label{vesicle shape evolution}\end{equation}
\begin{equation}
Q_{N}=\frac{Ca_{E}}{\lambda^{2}}\left[(\lambda-\alpha Q_{1}^{'}(\xi_{0}))^{2}+(\lambda-\alpha Q_{1}(\xi_{0})/\xi_{0})^{2}-2\beta^{2}/\epsilon_{r}\right],\end{equation}
\begin{equation}
Q_{T}=\frac{Ca_{E}}{\lambda^{2}}\left[(\lambda-\alpha Q_{1}^{'}(\xi_{0}))(\lambda-\alpha Q_{1}(\xi_{0})/\xi_{0})-\beta^{2}/\epsilon_{r}\right].\end{equation}
\end{subequations}
\end{widetext}
The functions $f_{14}(\xi_{0})$, $f_{15}(\xi_{0})$,
$f_{21}(\xi_{0})-f_{24}(\xi_{0})$, and $F$ are the same as those
used in Zhang \emph{et al.}, \cite{Zhang2012} and the detailed expressions are found
in the Appendix. $\epsilon_{r}\equiv\epsilon_{e}/\epsilon_{i}$ is the permittivity ratio. The factors $Q_{N}$ and $Q_{T}$ again arise
from the effects of the tangential and normal stresses, respectively.
$Ca_{E}\equiv r_{0}\epsilon_{e}E_{0}^{2}/\Gamma_{0}$ is the modified
electric capillary number. In the absence of electroporation, the
coefficients $\alpha$ and $\beta$ are given in Eqs. (\ref{a(t)})
and (\ref{b(t)}). Once the electroporation occurs, Eq. (\ref{electroporated ab})
is used instead. Similar to the droplet model, an examination of the
three terms in the numerator of Eq. (\ref{vesicle shape evolution})
reveals the contribution from the normal stress, tangential stress,
and membrane tension, respectively. The balance between these three
terms determines the equilibrium vesicle shape. The above equations are solved until the end of the pulse, $t=t_{p}$.

In the context of vesicle electrodeformation, the relaxation process
is equally important, and is more revealing of the underlying physical
processes. The governing equations are presented below. In the absence
of electroporation, Eq. (\ref{laplace for vesicle}) is solved
without an applied electric field. The resulting equation for $\xi_{0}$
remains the same as Eq. (\ref{vesicle shape evolution}). The
coefficients of $Q_{N}$, $Q_{T}$, $\alpha$, and $\beta$ are given
as\begin{equation}
Q_{N}=\frac{\epsilon_{e}V_{c}^{2}}{\lambda^{2}r_{0}\Gamma_{0}}\left[\alpha^{2}\left(Q_{1}^{'2}(\xi_{0})+Q_{1}^{2}(\xi_{0})/\xi_{0}^{2}\right)-2\beta^{2}/\epsilon_{r}\right],\end{equation}
\begin{equation}
Q_{T}=\frac{\epsilon_{e}V_{c}^{2}}{\lambda^{2}r_{0}\Gamma_{0}}\left[\alpha^{2}Q_{1}(\xi_{0})Q_{1}^{'}(\xi_{0})/\xi_{0}-\beta^{2}/\epsilon_{r}\right],\end{equation}
\begin{equation}
\alpha=\frac{\beta}{Q_{1}^{'}(\xi_{0})\sigma_{r}},\end{equation}
\begin{widetext}
\begin{equation}
\left[\frac{Q_{1}(\xi_{0})}{Q_{1}^{'}(\xi_{0})\sigma_{r}}-\xi_{0}\right]\frac{d\beta}{d\tau}-\left[\frac{Q_{1}(\xi_{0})Q_{1}^{''}(\xi_{0})-Q_{1}^{'2}(\xi_{0})(1-\sigma_{r})}{Q_{1}^{'2}(\xi_{0})\sigma_{r}}\frac{d\xi_{0}}{d\tau}+\frac{\tau_{2}}{\tau_{1}\lambda}\right]\beta=0,\end{equation}
\begin{equation}
\alpha(\tau_{p})=\frac{V_{m}(\tau_{p})}{V_{c}(Q_{1}^{'}(\xi_{0})\xi_{0}\sigma_{r}-Q_{1})},\qquad\beta(\tau_{p})=\frac{V_{m}(\tau_{p})Q_{1}^{'}(\xi_{0})\sigma_{r}}{V_{c}(Q_{1}^{'}(\xi_{0})\xi_{0}\sigma_{r}-Q_{1})}.\label{eq:intial relaxation}\end{equation}
\end{widetext}
In Eq. (\ref{eq:intial relaxation}), the initial conditions
for $\alpha$ and $\beta$ are obtained by solving Eqs. (\ref{laplace for vesicle})
and (\ref{eq:capacitive current continous}), and requiring that $V_{m}$
assumes the value at the end of the pulse. $\tau_{p}$ is the dimensionless
time, $t_{p}/\tau_{2}$. Note that in this case, although the pulse is switched off, the electric
field is in general not zero, due to the capacitive discharging of
the membrane. In this case, the TMP will decreases from its peak value
to zero on the membrane-charging timescale, $T_{ch}$.

When electroporation is present, the discharging process is slightly
more complex. The full membrane-charging model (\ref{eq:reduced current continous})
is used. In order to determine the membrane conductance, $G_{m}$,
we simply assume that it remains unchanged from the moment the pulse
ceases, namely, \begin{equation}
G_{m}=-\frac{\sigma_{e}\beta E_{0}}{\lambda V_{c}}.\end{equation}

The resulting equation for $\xi_{0}$ again does not formally deviate
from Eq. (\ref{vesicle shape evolution}). The coefficients of
$Q_{N}$, $Q_{T}$, $\alpha$, and $\beta$ are\begin{equation}
Q_{N}=\frac{\epsilon_{e}V_{c}^{2}}{\lambda^{2}r_{0}\Gamma_{0}}\left[\alpha^{2}\left(Q_{1}^{'2}(\xi_{0})+Q_{1}^{2}(\xi_{0})/\xi_{0}^{2}\right)-2\beta^{2}/\epsilon_{r}\right],\end{equation}
\begin{equation}
Q_{T}=\frac{\epsilon_{e}V_{c}^{2}}{\lambda^{2}r_{0}\Gamma_{0}}\left[\alpha^{2}Q_{1}(\xi_{0})Q_{1}^{'}(\xi_{0})/\xi_{0}-\beta^{2}/\epsilon_{r}\right],\end{equation}
\begin{equation}
\alpha=\frac{\beta}{Q_{1}^{'}(\xi_{0})\sigma_{r}},\end{equation}
\begin{widetext}
\begin{equation}
\left[\frac{Q_{1}(\xi_{0})}{Q_{1}^{'}(\xi_{0})\sigma_{r}}-\xi_{0}\right]\frac{d\beta}{d\tau}-\left[\frac{Q_{1}(\xi_{0})Q_{1}^{''}(\xi_{0})-Q_{1}^{'2}(\xi_{0})(1-\sigma_{r})}{Q_{1}^{'2}(\xi_{0})\sigma_{r}}\frac{d\xi_{0}}{d\tau}+\frac{\tau_{2}}{\tau_{1}\lambda}-\frac{\tau_{2}G_{m}}{C_{m}}\left(\frac{Q_{1}(\xi_{0})}{Q_{1}^{'}(\xi_{0})\sigma_{r}}-\xi_{0}\right)\right]\beta=0.\end{equation}
\end{widetext}

\subsection{A similarity solution for vesicle relaxation}

The governing equation for the relaxation process can be further simplified
following two considerations. First, we may ignore the membrane-discharging
process. The membrane-charging/discharging time, $T_{ch}$, is on
the order of 1 ms, which is in general much shorter than the relaxation
time observed in the experiments, namely, a few tens of ms or longer.
The relatively small effect of discharging on relaxation is clearly
seen in Fig. \ref{fig:effect of sigma0} presented in the following
section. Without including the discharging process, the coefficients
$Q_{T}$ and $Q_{N}$ in Eq. (\ref{vesicle shape evolution})
are simply set to zero. Second, in the membrane-mechanical model (\ref{membrane tension}),
the first and second term on the RHS represent the effects of undulation
unfolding and elastic stretching, respectively. For moderate values
of $\Gamma_{0}$, and for small-to-moderate deformations, the second
term can be ignored, and the membrane-mechanical model becomes\begin{equation}
\Delta=\frac{k_{B}T}{8\pi\kappa}{\rm ln}\frac{\Gamma_{h}}{\Gamma_{0}}.\label{eq:reduced membrane tension model}\end{equation}
Substituting $Q_{T}=Q_{N}=0$ and Eq. (\ref{eq:reduced membrane tension model})
into (\ref{vesicle shape evolution}), we obtain\begin{equation}
\frac{d\xi_{0}}{d\tau}=\frac{1}{F}{\rm exp}(\frac{8\pi\kappa\Delta}{k_{B}T})f_{24}(\xi_{0}).\label{eq:similarity}\end{equation}
This equation is conveniently rewritten in terms of the aspect ratio
as \begin{equation}
\frac{d\frac{a}{b}}{d\tau}=-\frac{1}{F}{\rm exp}(\frac{8\pi\kappa\Delta}{k_{B}T})(\xi_{0}^{2}-1)^{-\frac{3}{2}}f_{24}(\xi_{0}).\label{eq:reduced similarity}\end{equation}
Note that in this equation, $\kappa$, the bending rigidity, is regarded
constant for a specific vesicle type, and $\mu_{r}$ (embedded in
$F$, see Appendix) is close to 1 as both the
fluids are usually aqueous. In addition, $\Delta$, the relative increase
of apparent membrane area, depends exclusively on $\xi_{0}$, hence
$a/b$ according to Eqs. (\ref{eq:relative area increase}) and
(\ref{ab}). Under these assumptions, we observe
that Eq. (\ref{eq:reduced similarity}) is completely autonomous,
and the relaxation process is governed by the dimensionless time,
$\tau=t/\tau_{2}$, where $\tau_{2}=r_{0}\mu_{e}/\Gamma_{0}$. This
result suggests that the relaxation of vesicles with different initial
radius, $r_{0}$, and initial tension, $\Gamma_{0}$, obeys a similarity
behavior with the proper scaling suggested above. This behavior is
demonstrated by both simulation and analysis of previous experimental
data below.

\section{Results}

For all results below, we assume the lipid membrane to be made of
egg-PC following Riske and Dimova\cite{Riske2005} (henceforth abbreviated as $'\rm{RD05}'$)
and Sadik \emph{et al.}\cite{Sadik2011} (henceforth denoted as $'\rm{S11}'$). The bending
rigidity is taken to be $\kappa=2.47\times10^{-20}\;{\rm J}$; \cite{Kummrow1991}
the elastic modulus, $K_{a}=0.14\;{\rm N/m}$;\cite{Kwok1981,Needham1995} the membrane capacitance, $C_{m}=0.01\;{\rm F/m^{2}}$;
\cite{Needham1989} the intravesicular and extravesicular viscosities,
$\mu_{i}=\mu_{e}=10^{-3}\;{\rm Pa\cdot s}$; the intravesicular and
extravesicular permittivities, $\epsilon_{i}=\epsilon_{e}=7\times10^{-10}\:{\rm F/m}$.
The critical transmembrane potential is assumed to be $V_{c}=1\:{\rm V}$.
\cite{Portet2010}

\subsection{The effects of $\Gamma_{0}$ and $t_{p}$}

\begin{figure*}
\center

\includegraphics[width=0.5\textwidth]{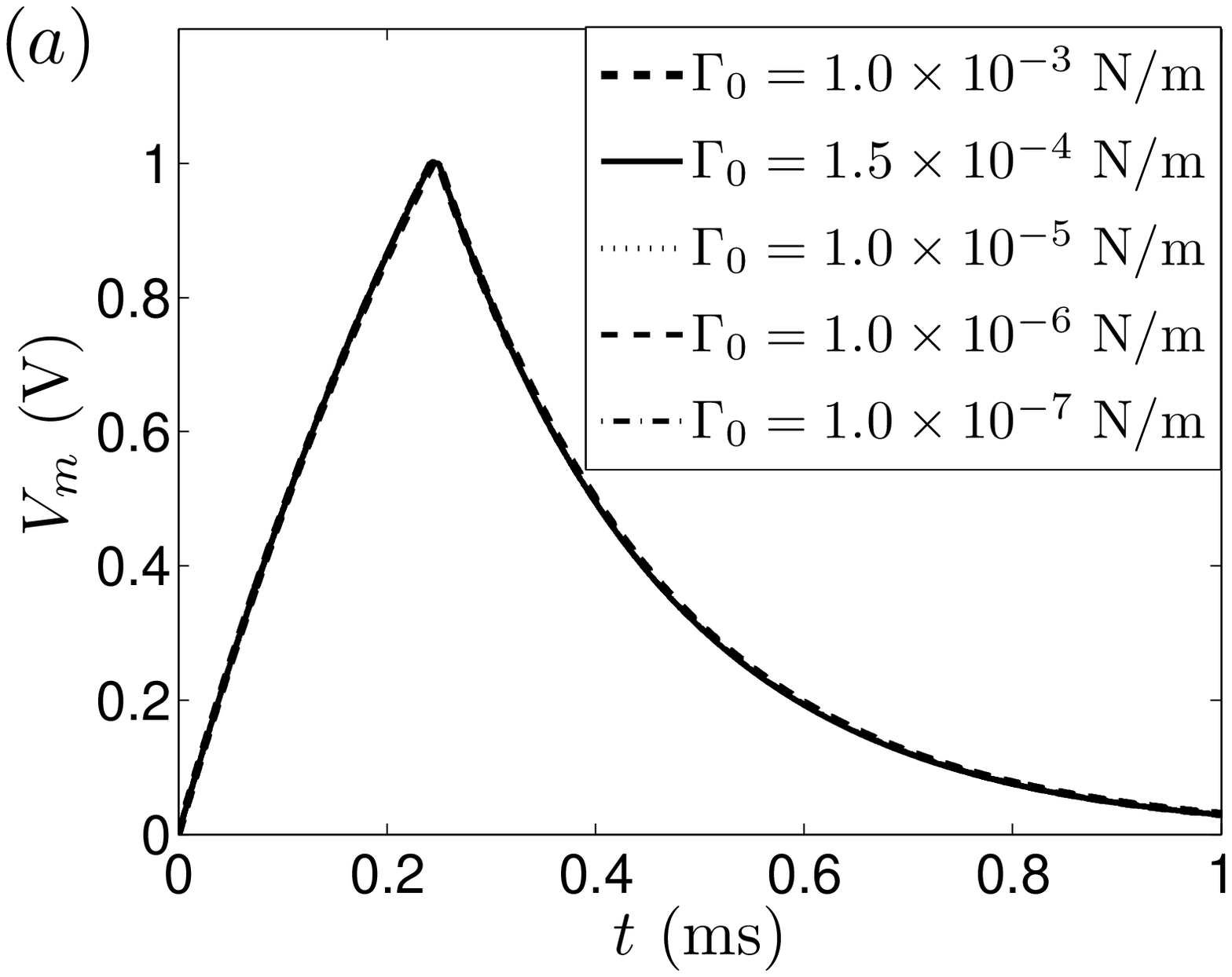}\includegraphics[width=0.5\textwidth]{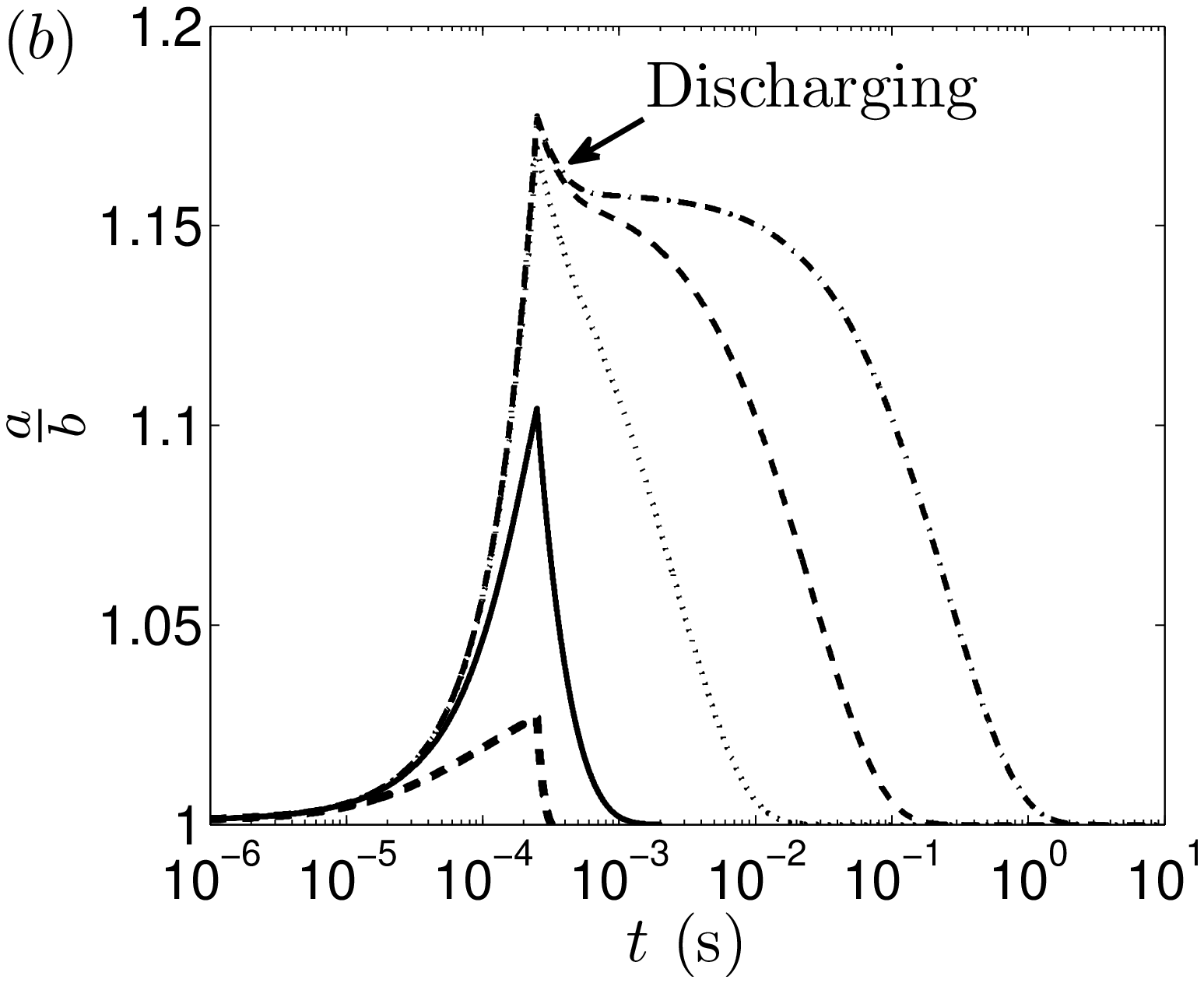}

\includegraphics[width=0.5\textwidth]{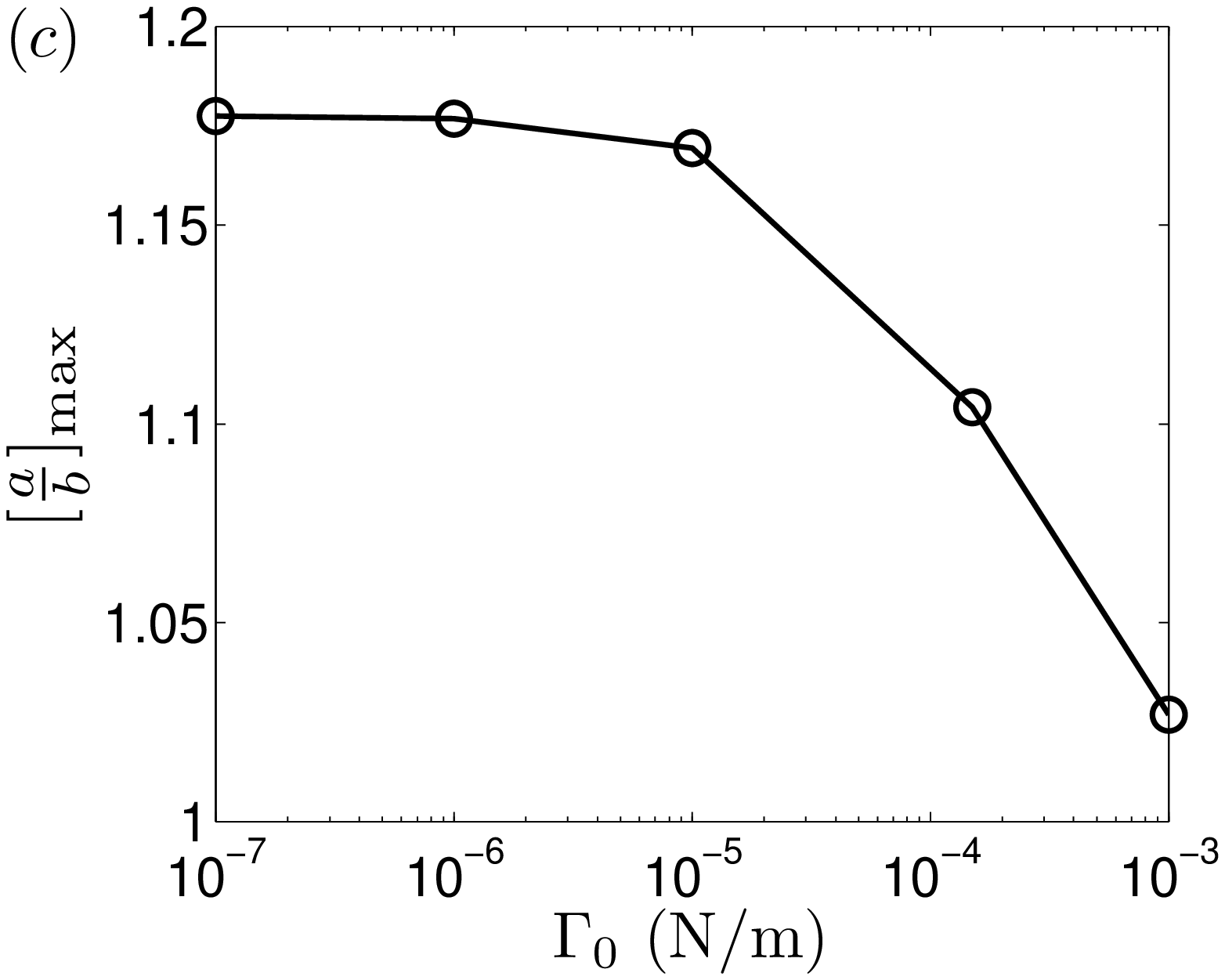}\includegraphics[width=0.5\textwidth]{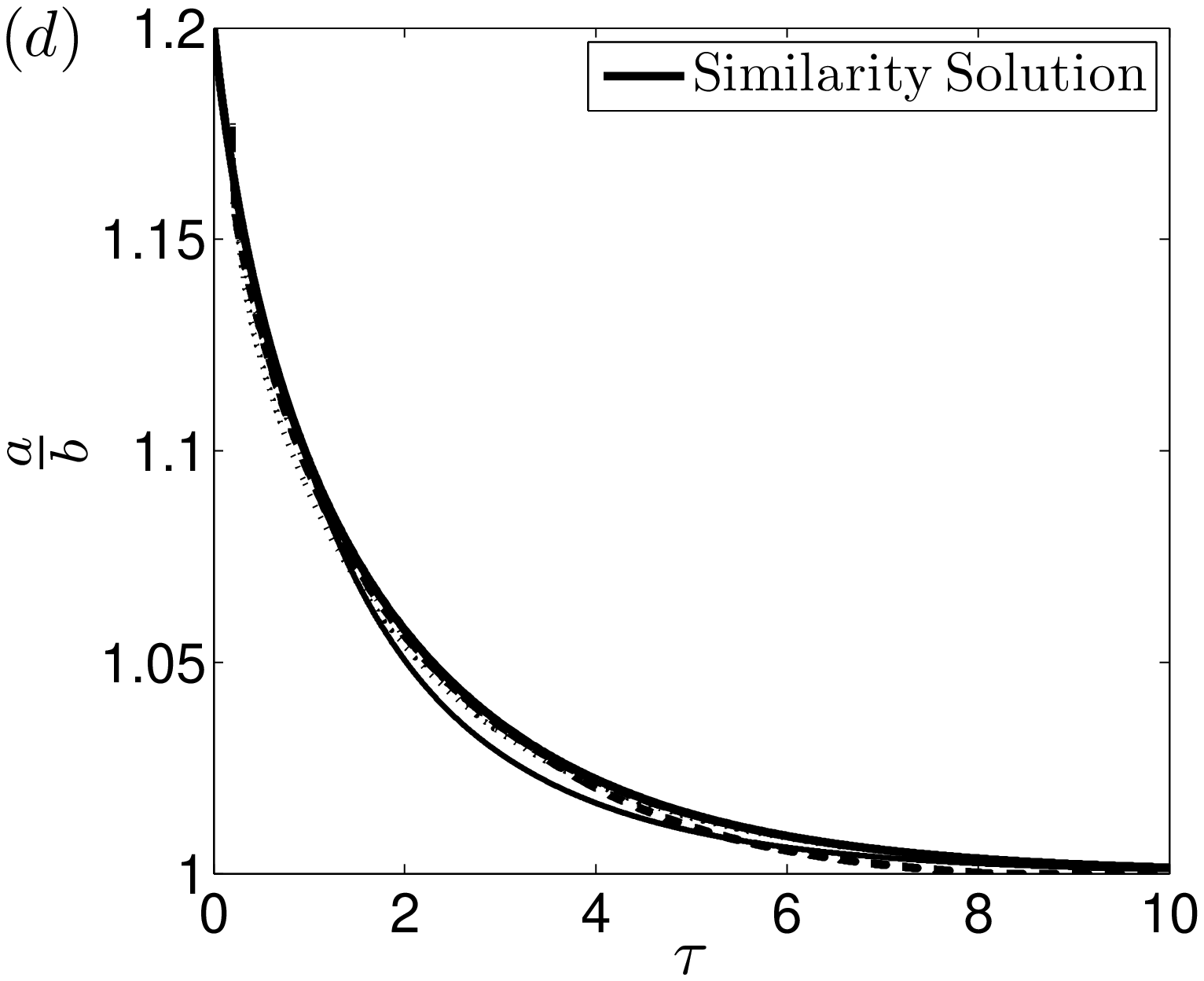}\caption{Vesicle deformation-relaxation as a function of $\Gamma_{0}$. The
governing parameters are $\sigma_{i}=6\times10^{-4}\:{\rm S/m}$,
$\sigma_{e}=4.5\times10^{-4}\:{\rm S/m}$, \emph{$E_{0}=1\:{\rm kV/cm}$},
$t_{p}=250\:{\rm \mu s}$, and $r_{0}=15\:{\rm \mu m}$. (a) The transmembrane
potential at the cathode-facing pole. (b) The time-course of the aspect
ratio. (c) The maximum aspect ratio as a function of $\Gamma_{0}$.
(d) The similarity behavior in relaxation. The descending branches
from (b) are rescaled with $\tau=t/\tau_{2}$. The thick solid curve
is directly obtained by integrating Eq. (\ref{eq:reduced similarity}).\label{fig:effect of sigma0}}

\end{figure*}

We begin by examining the effects of $\Gamma_{0}$ on vesicle electrodeformation
and relaxation. Figure \ref{fig:effect of sigma0} shows the typical
system behavior for values of $\Gamma_{0}$ ranging from $10^{-7}-10^{-3}\;{\rm N/m}$.
The intravesicular and extravesicular conductivities are $\sigma_{i}=6\times10^{-4}\;{\rm S/m}$
and $\sigma_{e}=4.5\times10^{-4}\;{\rm S/m}$, respectively following
RD05. The field strength is \emph{$E_{0}=1\;{\rm kV/cm}$}, the pulse
length is $t_{p}=250\;{\rm \mu s}$, and the initial radius is $r_{0}=15\;{\rm \mu m}$.
Figure \ref{fig:effect of sigma0}(a) shows the evolution of $V_{m}$
at the cathode-facing pole, which demonstrates only a weak dependence
on $\Gamma_{0}$. The threshold for electroporation ($1\:{\rm V}$)
is reached just before the end of the pulse, and its effects are present
yet negligible. The discharging occurs on the relatively short timescale
of $1\;{\rm ms}$ as we discussed above. Figure \ref{fig:effect of sigma0}(b)
shows the evolution of the aspect ratio, $a/b$. The discharging process
manifests itself as a sudden and slight decrease in the aspect ratio
immediately after the pulse ceases; its effects can in general be
ignored without significantly altering the relaxation behavior. A
smaller value of $\Gamma_{0}$ leads to a larger aspect ratio, and
a longer relaxation process. The maximum aspect ratio, $[a/b]_{{\rm max}}$,
is plotted as a function of $\Gamma_{0}$ in Fig. \ref{fig:effect of sigma0}(c).
As the initial membrane tension decreases toward zero, the maximum
achievable aspect ratio saturates. 

The similarity behavior in the relaxation process is demonstrated
in Fig. \ref{fig:effect of sigma0}(d). The descending branches of
the curves ($t>t_{p}$) shown in Fig. \ref{fig:effect of sigma0}(b)
are rescaled in terms of $\tau=t/\tau_{2}$, and shifted horizontally.
In comparison, the thick solid curve is obtained by directly solving
Eq. (\ref{eq:reduced similarity}). The convergence of all curves
validates that $\tau_{2}=r_{0}\mu_{e}/\Gamma_{0}$ is the single timescale
governing vesicle relaxation.%

\begin{figure}
\center

\includegraphics[width=0.5\textwidth]{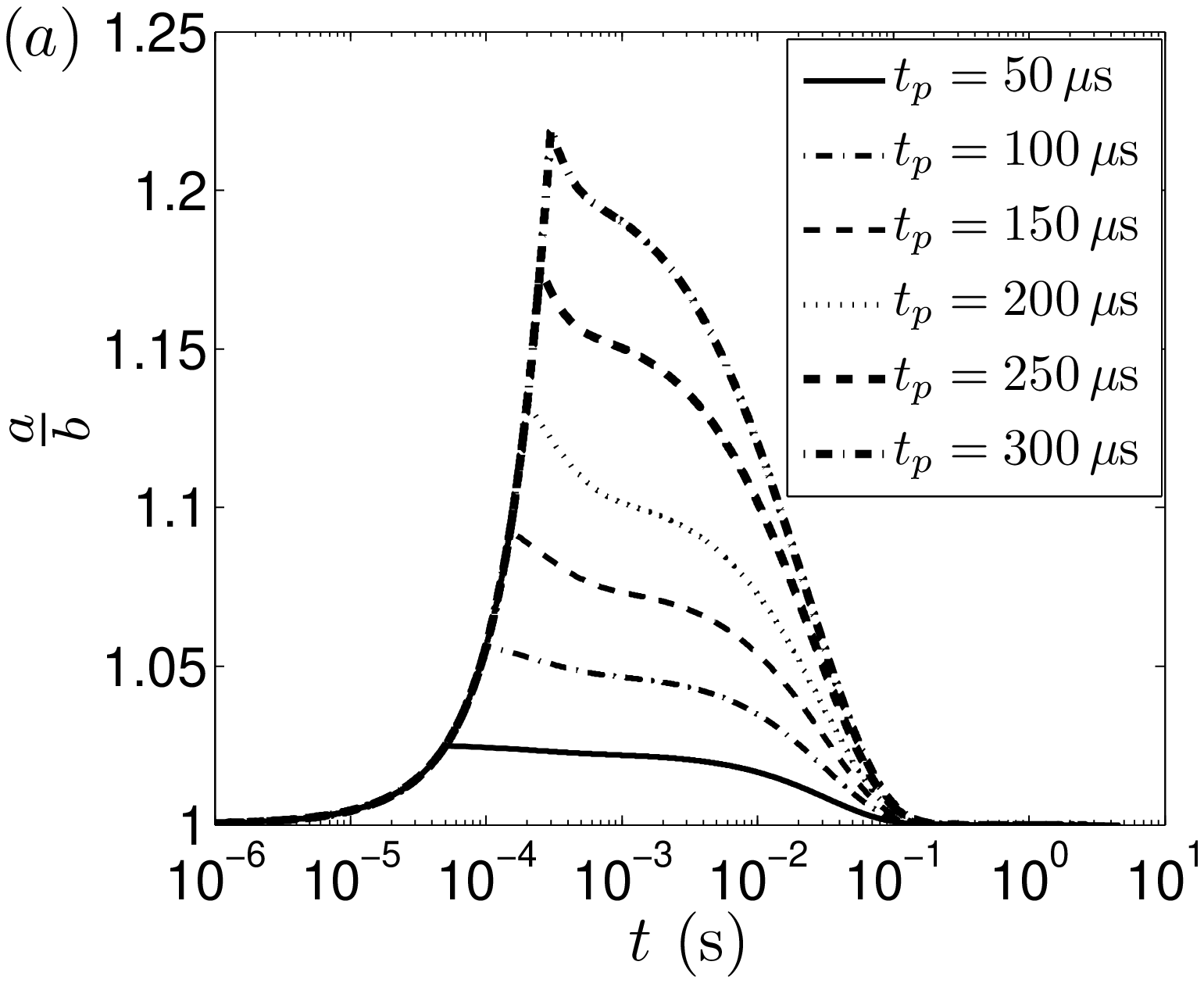}
\includegraphics[width=0.5\textwidth]{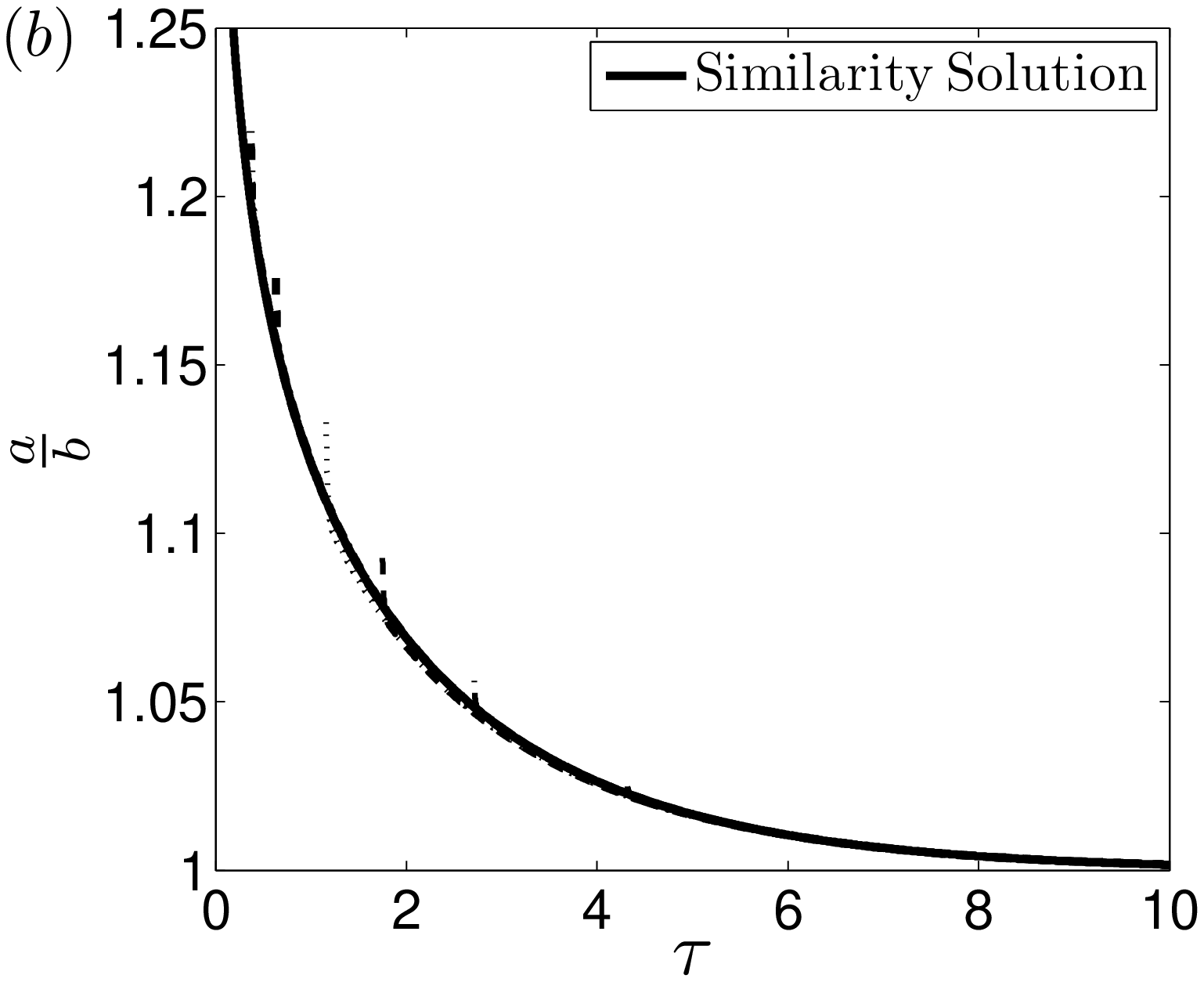}\caption{Vesicle deformation-relaxation as a function of $t_{p}$. The parameters
are the same as in Fig. \ref{fig:effect of sigma0}. The initial
tension is set to be constant, $\Gamma_{0}=1\times10^{-6}\:{\rm N/m}$.
(a) The time-course of the aspect ratio. (b) The similarity behavior
is observed by shifting the relaxation curves with respect to time.
The relaxation timescale, $\tau_{2}=r_{0}\mu_{e}/\Gamma_{0}$, is
the same for all cases. The thick solid curve is directly obtained
by integrating Eq. (\ref{eq:reduced similarity}).\label{fig:effect of tp}}

\end{figure}

\begin{table}
\caption{List of parameters for Fig. \ref{fig:Comparison with RD05}. For
each case, $E_{0}$ and $t_{p}$ are specified according to RD05.
$\Gamma_{0}$ is a fitting parameter to obtain best comparison between
simulation and data. For cases b, d, e, and f, extended pulse lengths
(denoted by star) are also used. \label{tab:Listed-parameters}}
\begin{ruledtabular}
\center\begin{tabular}{>{\centering}p{1cm}>{\centering}p{1cm}>{\centering}p{2cm}>{\centering}p{3cm}}
\multicolumn{1}{c}{case \#}  &\multicolumn{1}{c}{$E_{0}$ (kV/cm)}  &\multicolumn{1}{c}{$t_{p}$ (${\rm \mu s}$)} &\multicolumn{1}{c}{$\Gamma_{0}$ (N/m)}\tabularnewline
\hline
a & 1 & 150 & $2.79\times10^{-4}$\tabularnewline
b & 1 & 200 & $3.23\times10^{-6}$\tabularnewline
 & 1 & 300{*} & $3.23\times10^{-6}$\tabularnewline
c & 1 & 250 & $1.67\times10^{-4}$\tabularnewline
d & 1 & 300 & $1.80\times10^{-6}$\tabularnewline
 & 1 & 400{*} & $1.80\times10^{-6}$\tabularnewline
e & 2 & 50 & $1.80\times10^{-4}$\tabularnewline
 & 2 & 80{*} & $1.80\times10^{-4}$\tabularnewline
f & 2 & 100 & $3.16\times10^{-6}$\tabularnewline
 & 2 & 170{*} & $3.16\times10^{-6}$\tabularnewline
g & 3 & 50 & $6.67\times10^{-6}$\tabularnewline
h & 3 & 100 & $3.42\times10^{-7}$\tabularnewline
\end{tabular}
\end{ruledtabular}

\end{table}

The effects of $t_{p}$ are examined in Fig. \ref{fig:effect of tp}.
The parameters are the same as in Fig. \ref{fig:effect of sigma0},
and we fix $\Gamma_{0}$ at $1\times10^{-6}\;{\rm N/m}$. Figure \ref{fig:effect of tp}(a)
shows that a longer pulse consistently leads to greater deformation,
and the aspect ratio increases along the same envelope. The relaxation
times are approximately the same for all cases, because $\tau_{2}$
remains unchanged. The discharging process is in general more conspicuous
with longer pulses. In Fig. \ref{fig:effect of tp}(b), the relaxation
curves are again shifted horizontally and rescaled with $\tau_{2}$
to show good agreement with the similarity solution (thick solid line).
Note that here because all cases share the same values of $\tau_{2}$,
the collapse of the curves is primarily caused by simple shifting.
In other words, the aspect ratio also decreases along a common envelope.

The above results are exemplary and demonstrate the typical system
behavior. In general, the relaxation process (in particular the relaxation
time) is more appreciably affected by the change in $\Gamma_{0}$
than the deformation process. A wide range of pulsing parameters are
studied below, in direct comparison with experimental data from RD05
and S11.

\begin{figure*}
\center

\includegraphics[trim=0cm 1cm 0cm 0.5cm, clip=true, width=0.45\textwidth]{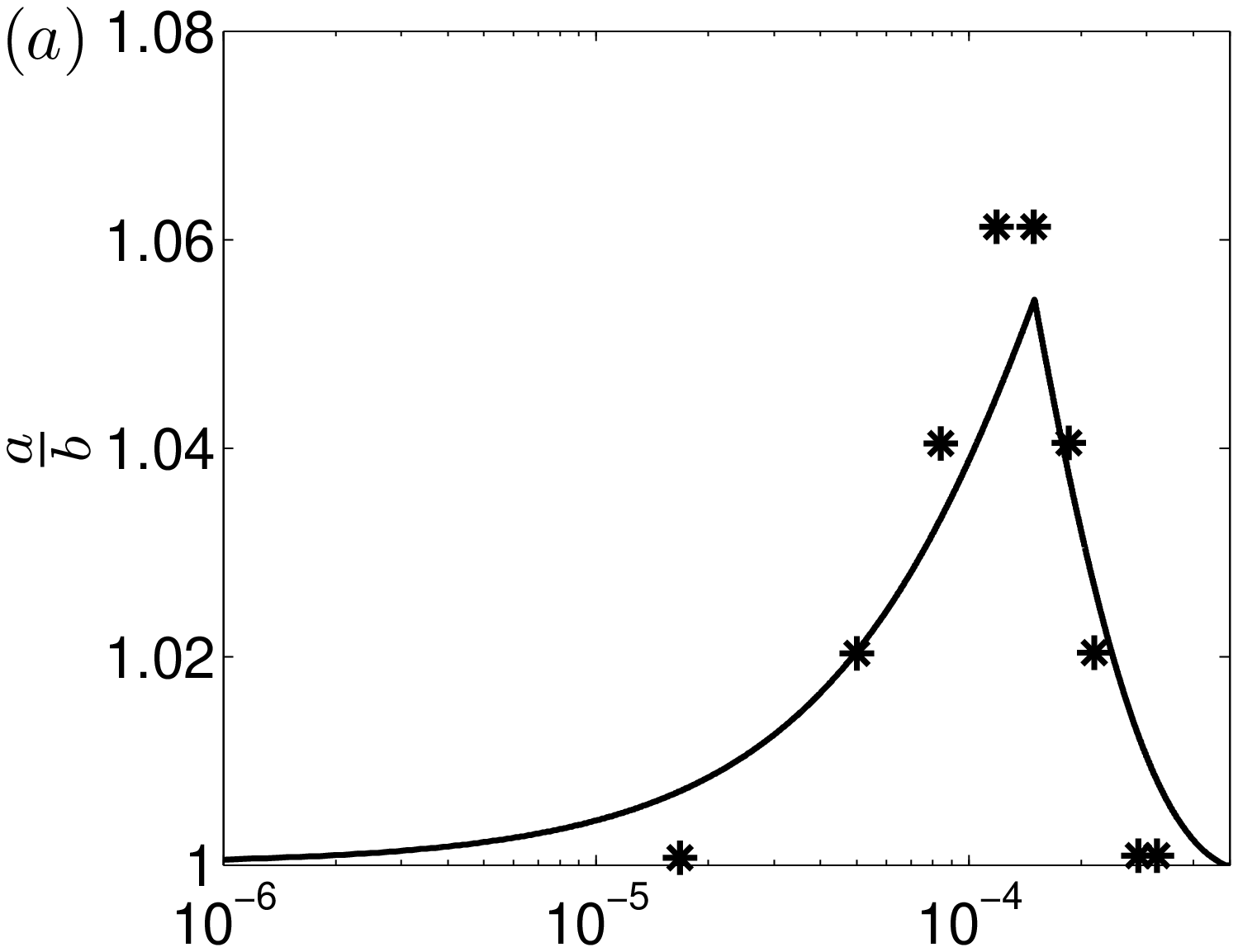}\includegraphics[trim=0cm 1cm 0cm 0.5cm, clip=true,width=0.45\textwidth]{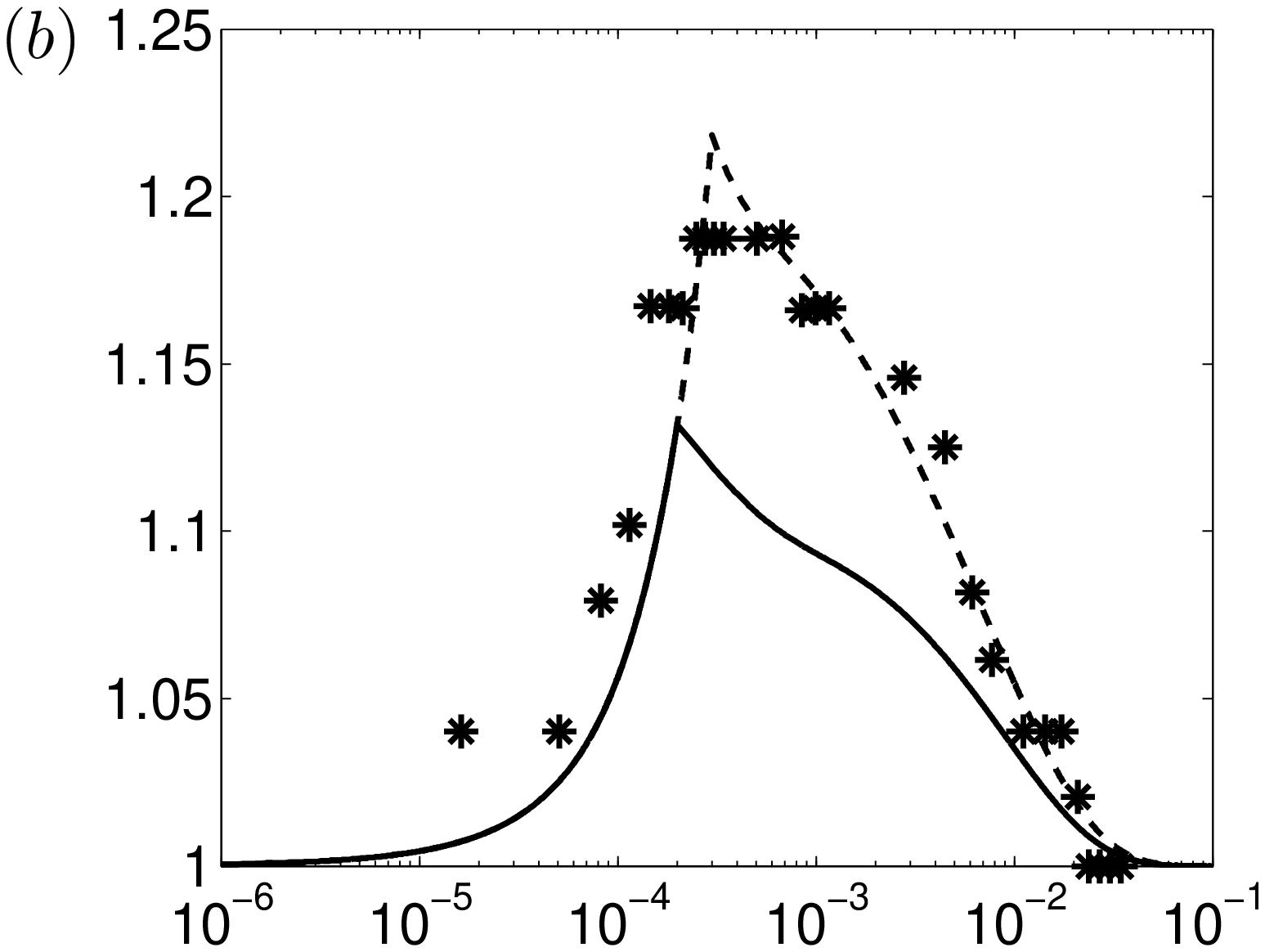}

\includegraphics[trim=0cm 1cm 0cm 0.5cm, clip=true,width=0.45\textwidth]{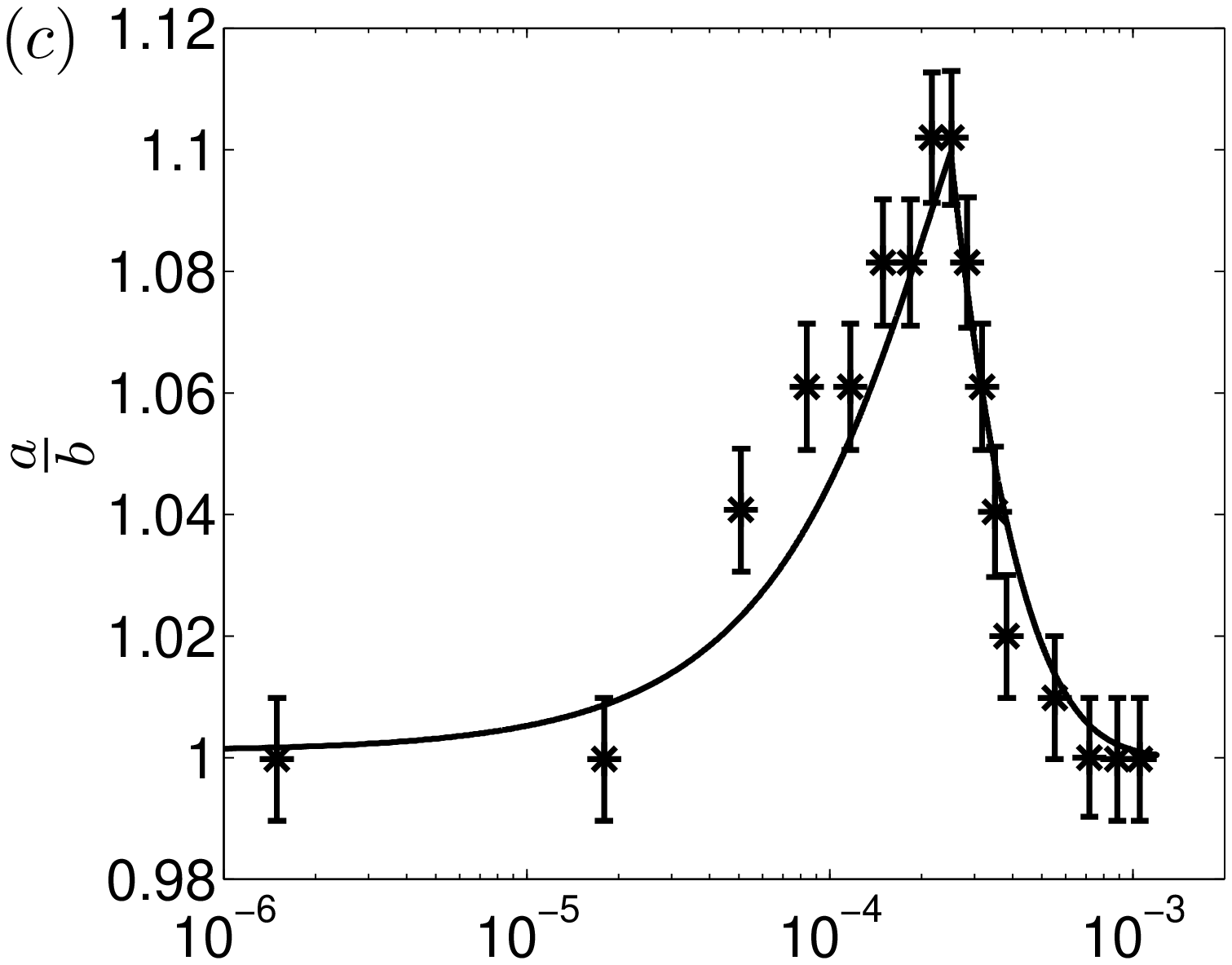}\includegraphics[trim=0cm 1cm 0cm 0.5cm, clip=true,width=0.45\textwidth]{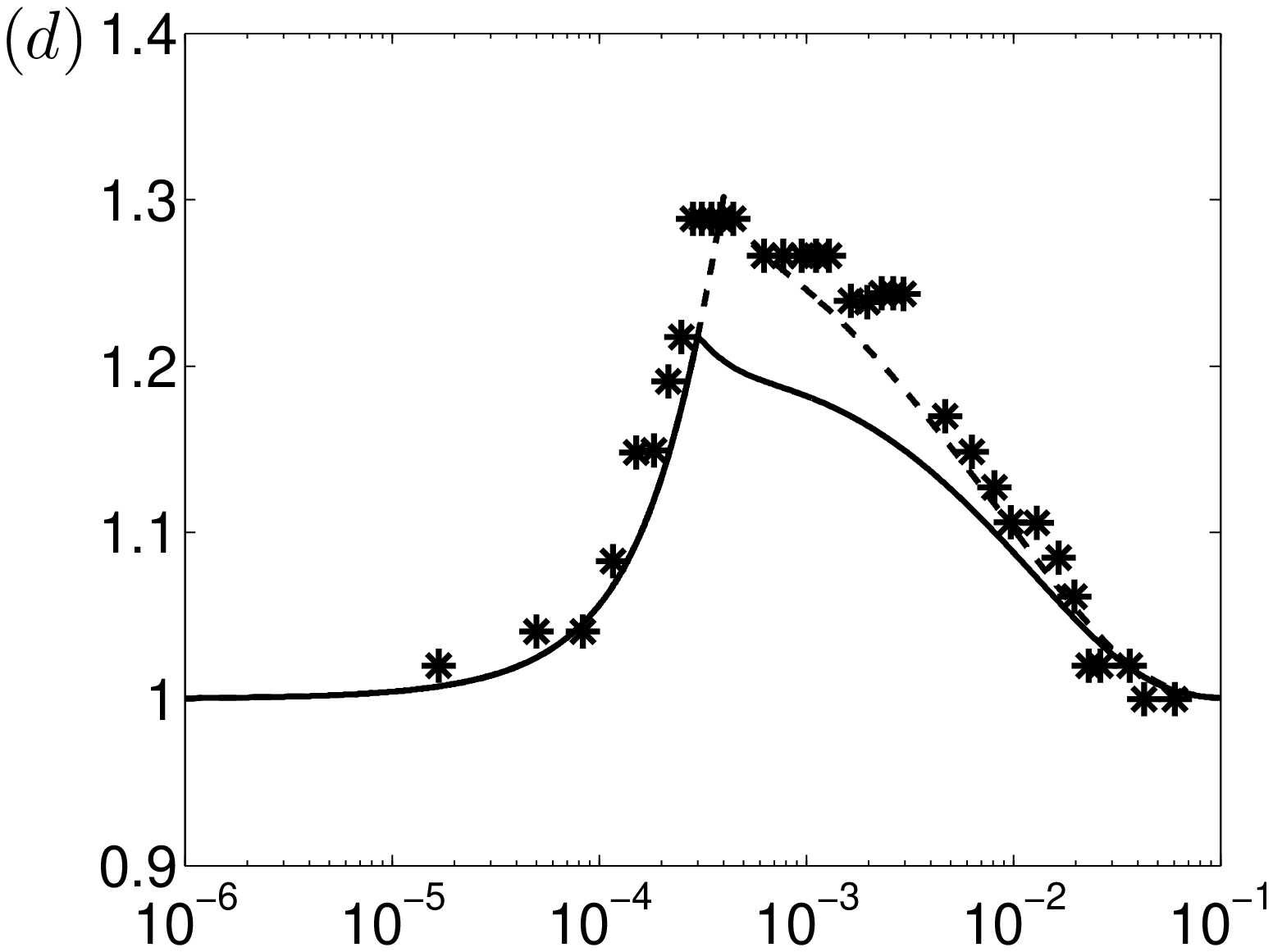}

\includegraphics[trim=0cm 1cm 0cm 0.5cm, clip=true,width=0.45\textwidth]{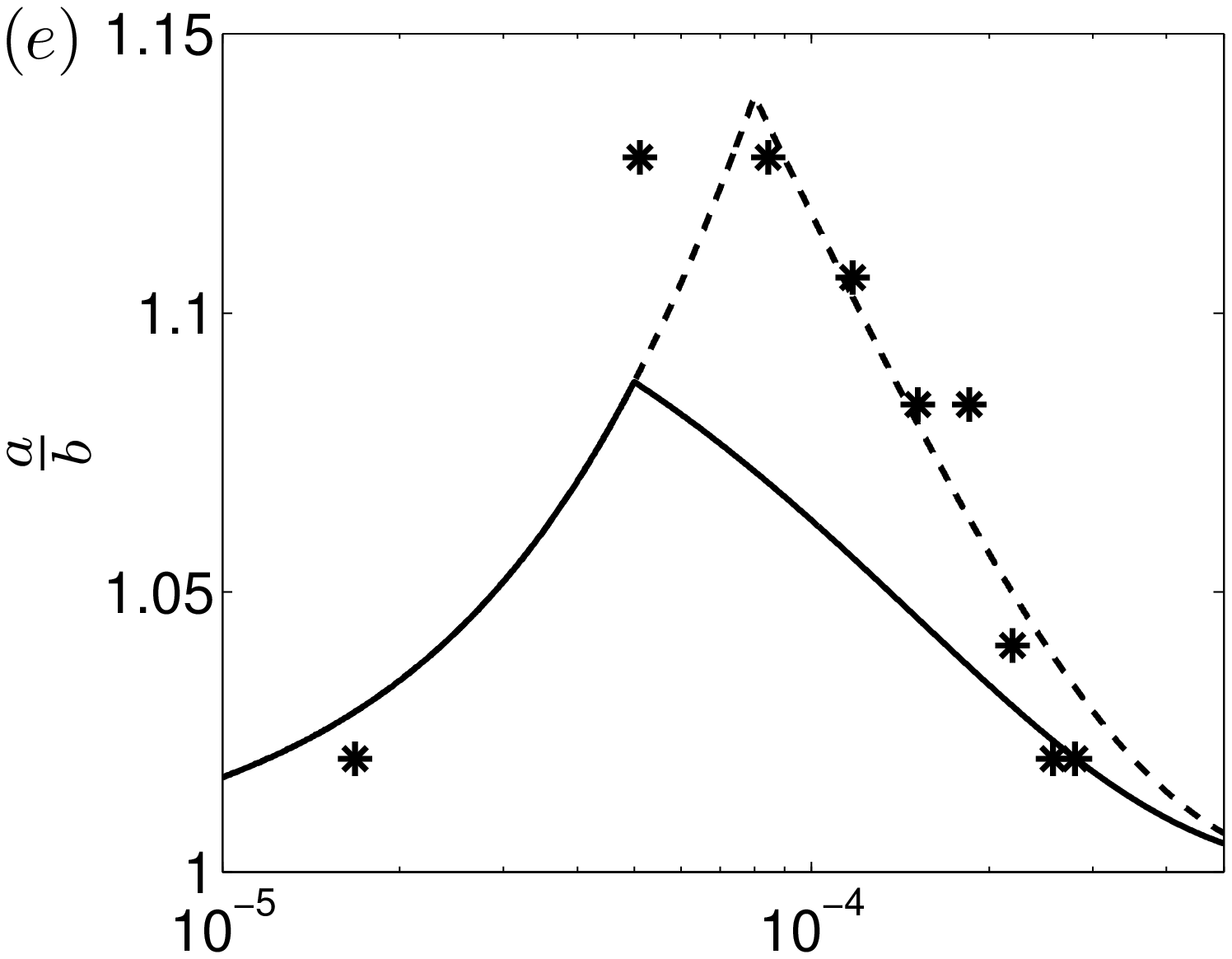}\includegraphics[trim=0cm 1cm 0cm 0.5cm, clip=true,width=0.45\textwidth]{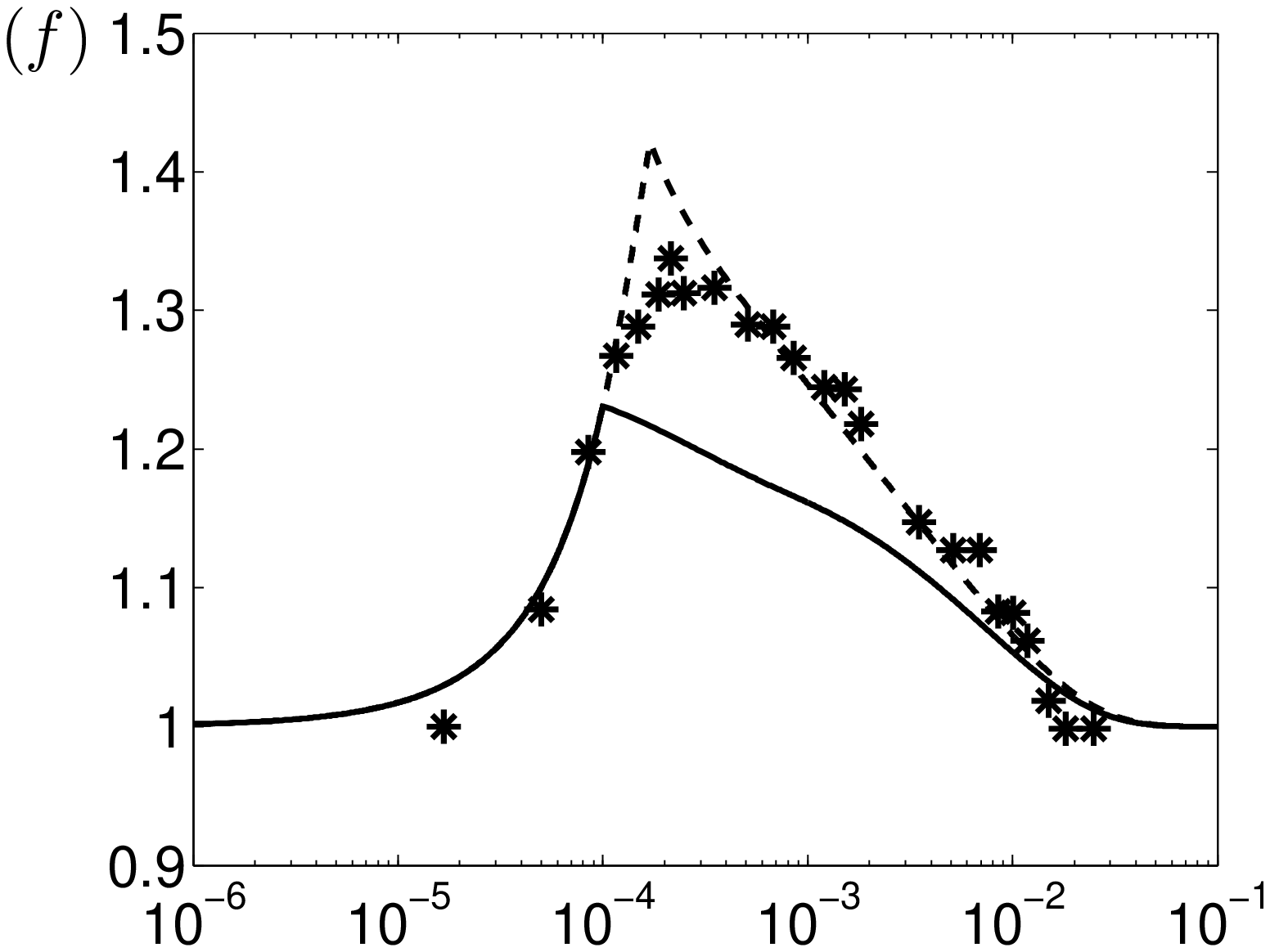}

\includegraphics[trim=0cm 0.2cm 0cm 0.5cm, clip=true,width=0.45\textwidth]{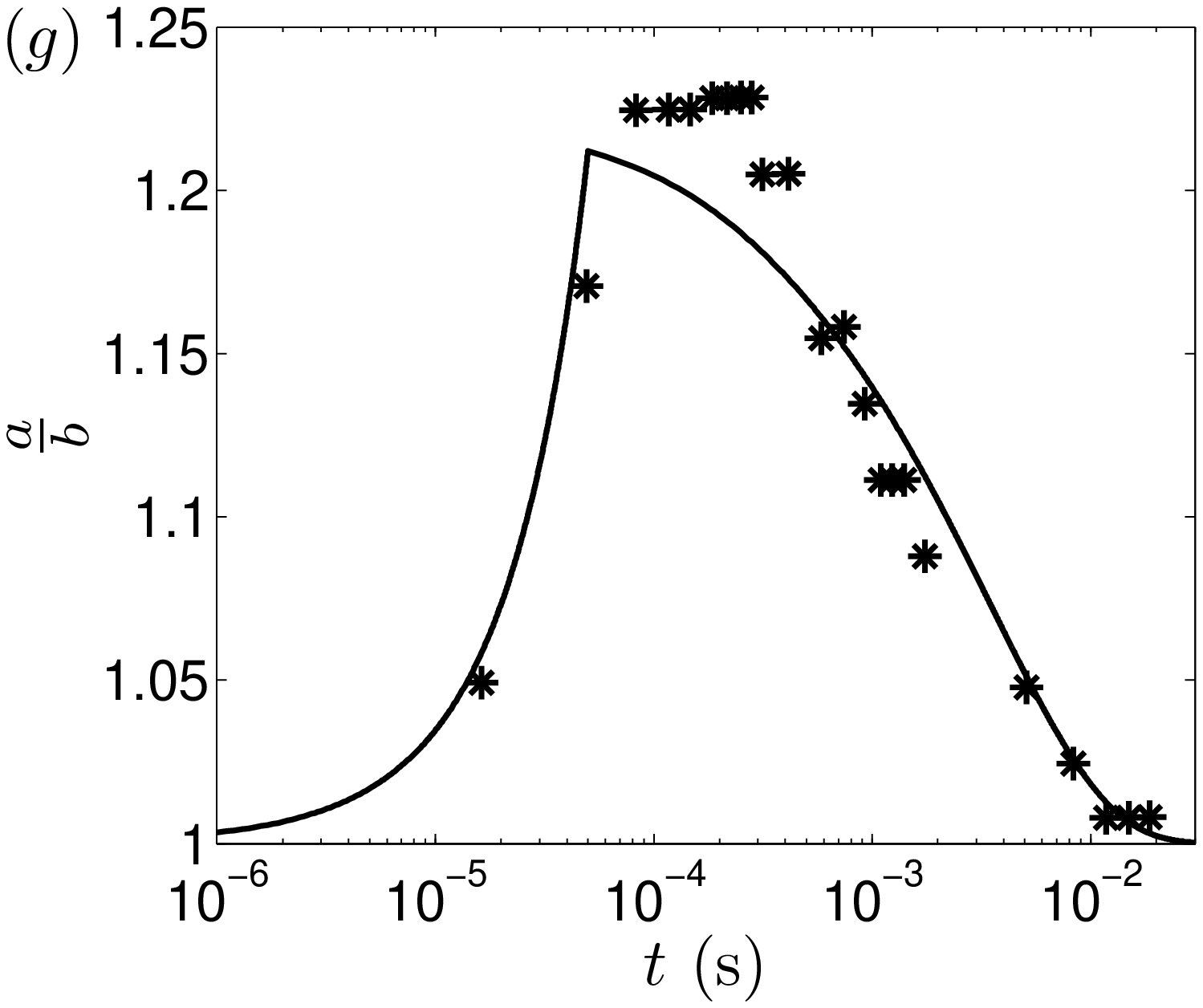}\includegraphics[trim=0cm 0.2cm 0cm 0.5cm, clip=true,width=0.45\textwidth]{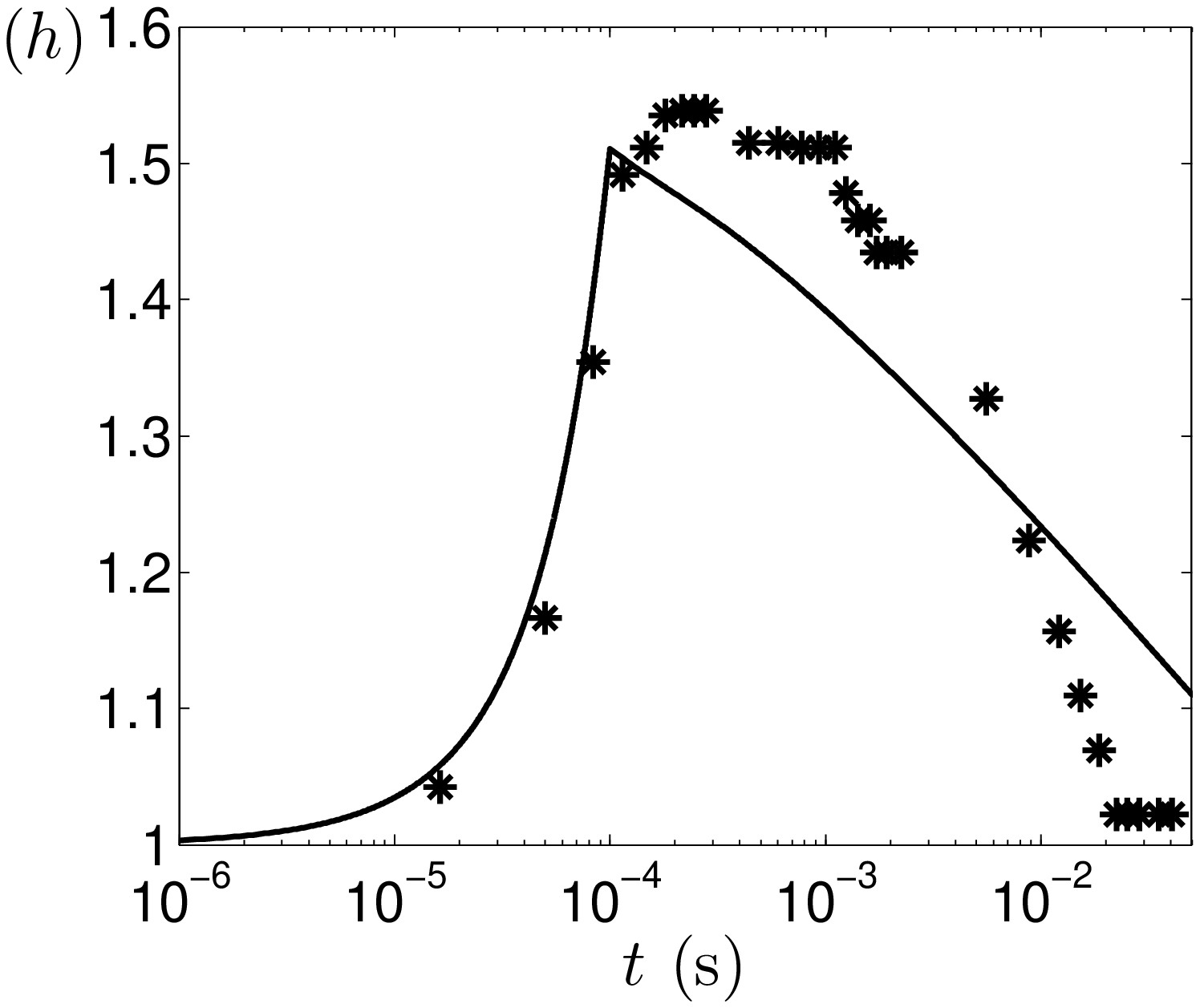}

\caption{Comparison with the deformation-relaxation data from RD05. For all
cases, $r_{0}=15\:{\rm \mu m}$, $\sigma_{i}=6\times10^{-4}\:{\rm S/m}$,
and $\sigma_{e}=4.5\times10^{-4}\:{\rm S/m}$. Parameters specific
to each case are listed in table \ref{tab:Listed-parameters}. The
data is represented by symbols, and the simulation is represented
by solid curves. For cases b, d, e, and f, the dashed lines represent
the simulated results with extended pulses (denoted by stars in table
\ref{tab:Listed-parameters}).\label{fig:Comparison with RD05}}

\end{figure*}

\subsection{Comparison with experimental data}

An extensive comparison of our theoretical prediction with the data
from RD05 is presented in Fig. \ref{fig:Comparison with RD05}.
For all eight cases, the initial radius is $r_{0}=15\;{\rm \mu m}$.
The electrical conductivities are $\sigma_{i}=6\times10^{-4}\;{\rm S/m}$
and $\sigma_{e}=4.5\times10^{-4}\;{\rm S/m}$, respectively, leading
to a conductivity ratio of $\sigma_{r}=0.75$. Other parameters are
listed in table \ref{tab:Listed-parameters}. All parameters are taken
directly from RD05, except for the extended pulse lengths for some
cases noted below. For each case, the initial tension, $\Gamma_{0}$,
is determined to best fit the experimental data; their values are
listed in table \ref{tab:Listed-parameters} in the last column. The
experimental data are presented as symbols; the theoretical predictions,
solid lines. In Figs. \ref{fig:Comparison with RD05}(a) to \ref{fig:Comparison with RD05}(d),
the electric field strength is $E_{0}=1\;{\rm kV/cm}$. For these
cases, $V_{m}$ is predicted to reach $V_{c}$ at $t=242\;{\rm \mu s}$.
In Figs. \ref{fig:Comparison with RD05}(a) and \ref{fig:Comparison with RD05}(c),
good agreements are observed between the theoretical prediction and
the data. In Figs. \ref{fig:Comparison with RD05}(b) and \ref{fig:Comparison with RD05}(d),
the model results underpredict the maximum aspect ratios. This discrepancy
is peculiar: our simulation follows the data accurately during the
presence of the pulse, which duration is provided by RD05. After the
pulse ceases, the simulation predicts immediate relaxation, whereas
the vesicles continued to deform in the experiments, due to some unknown
cause. In an attempt to mend this difference, we artificially increase
the pulse lengths in the simulation in b and d from 200 and 300 to
300 and 400 ${\rm \mu s}$, respectively. The values for $\Gamma_{0}$
remain unchanged. The results are shown as dashed curves. The model
predicts well the data for both the deformation and relaxation processes.
Note that although the relaxation curves represented by the solid
and dashed lines look somewhat different due to the semi-log scale
on the time axis, they actually follow the same descending envelopes
which we have demonstrated in Fig. \ref{fig:effect of tp}(b) above.

In Figs. \ref{fig:Comparison with RD05}(e) and \ref{fig:Comparison with RD05}(f),
the field strength is increased to be $E_{0}=2\;{\rm kV/cm}$, and
the pulse lengths used in RD05 were 50 and 100 ${\rm \mu s}$, respectively.
For these cases, our model predicts the occurrence of electroporation
around $t=103\;{\rm \mu s}$. A similar situation is observed as in
Figs. \ref{fig:Comparison with RD05}(b) and \ref{fig:Comparison with RD05}(d). The solid curves underpredict the maximum aspect ratio. Artificially
extending the pulses in e and f to 80 and 170 ${\rm \mu s}$, respectively,
leads to much better agreement between the two. 

In Figs. \ref{fig:Comparison with RD05}(g) and \ref{fig:Comparison with RD05}(h),
the field strength is further increased to 3 kV/cm, and electroporation
is predicted to occur at $t=66\;{\rm \mu s}$. The entire deformation-relaxation
process is well-captured in g where $t_{p}=50\;{\rm \mu s}$. In Fig.
\ref{fig:Comparison with RD05}(h), where $t_{p}=100\;{\rm \mu s}$,
although the model accurately predicts the deformation, the simulated
relaxation curve completely deviates from the experimental data. For
this case, and for pulses even longer than 100 ${\rm \mu s}$, RD05
[Fig. 1(c) therein] exhibits a regime where complex, multi-stage relaxation
process was observed. In this regime, the membrane structure is likely
severely altered due to electroporation, which process can not be
captured by our present model. Further comparison with these data
is not pursued.

\begin{figure}
\center\includegraphics[width=0.5\textwidth]{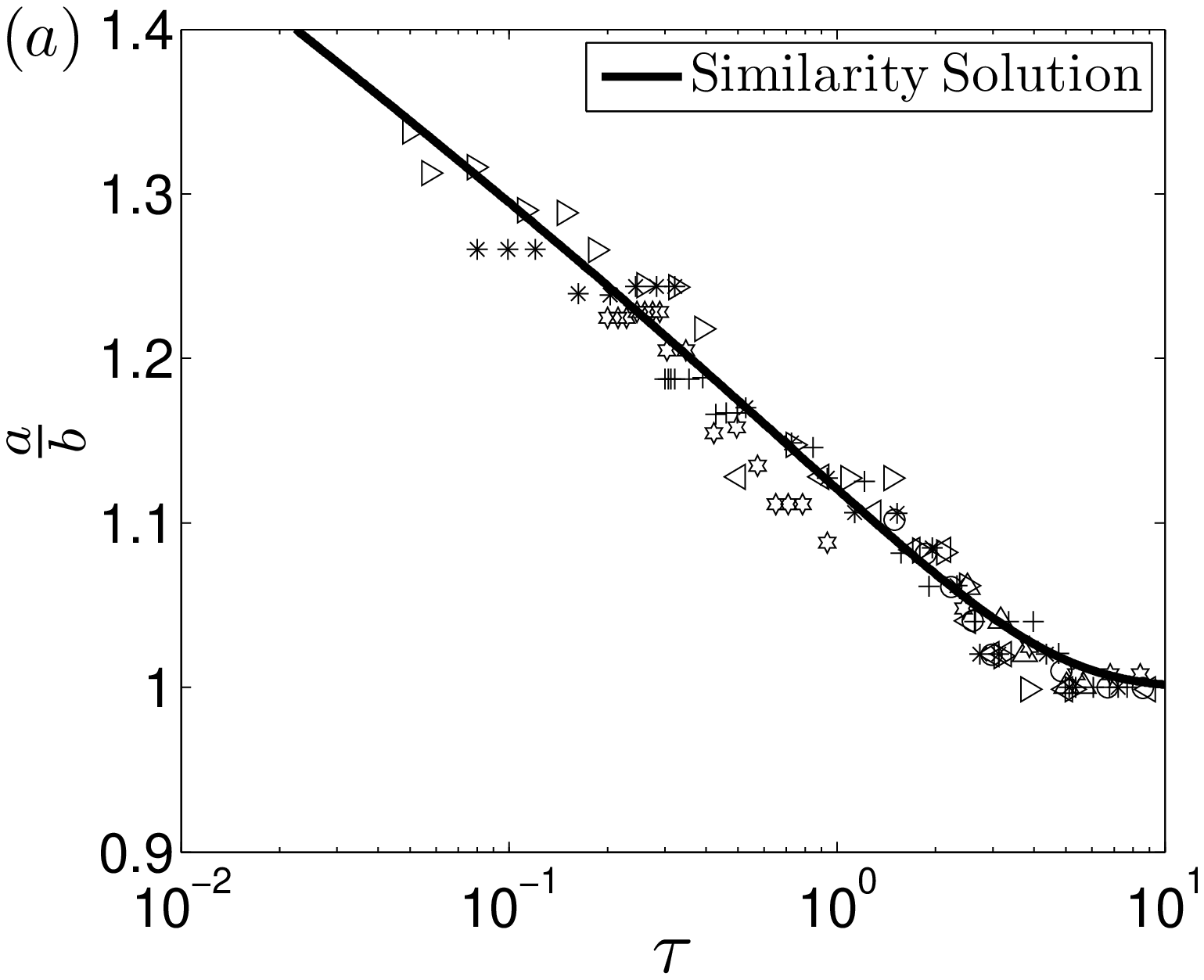}
\includegraphics[width=0.5\textwidth]{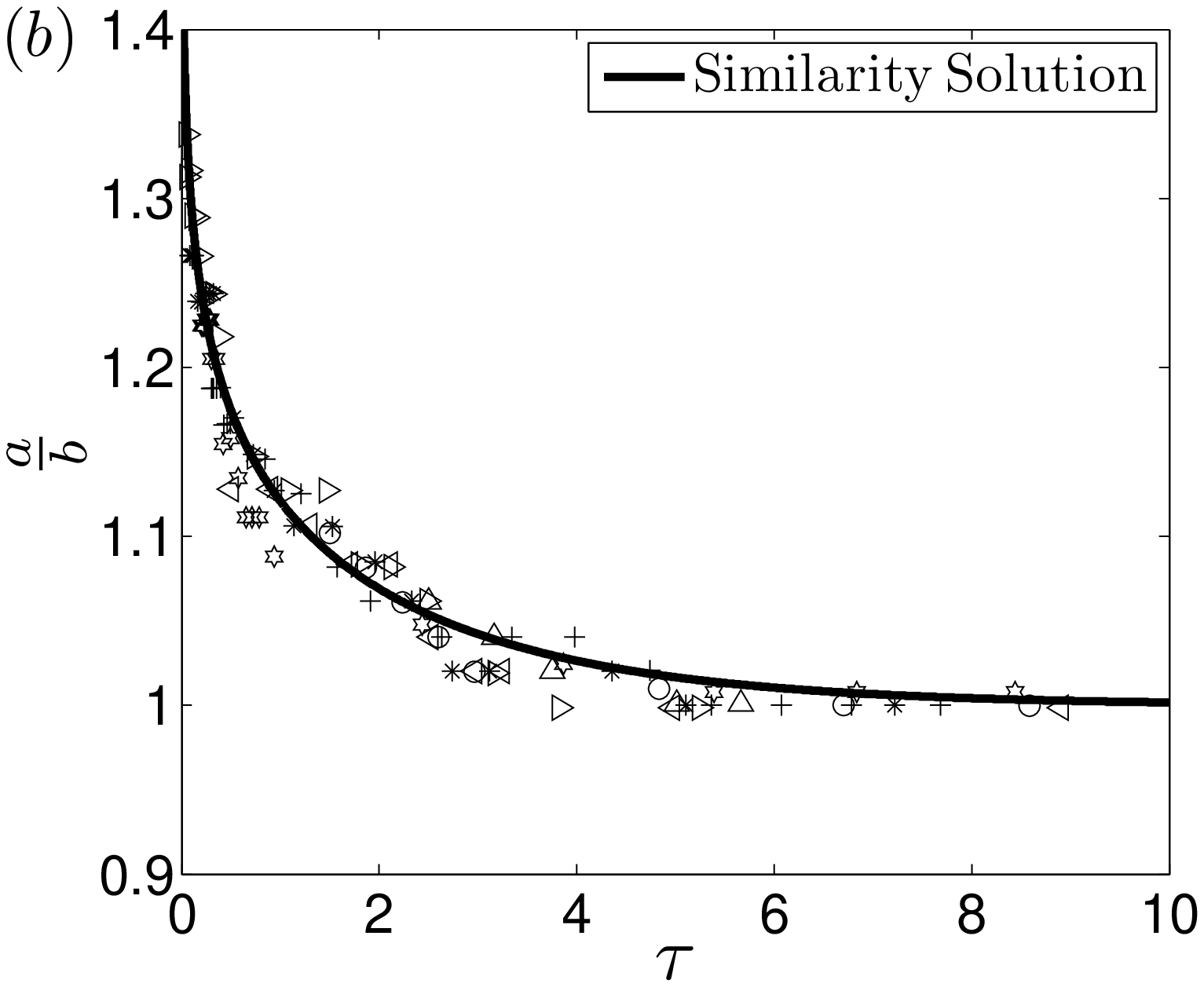}

\caption{The similarity behavior of vesicle relaxation. The experimental data
from cases a-g in Fig. \ref{fig:Comparison with RD05} are shifted
in time, then rescaled by $\tau_{2}=r_{0}\mu_{e}/\Gamma_{0}$. They
are represented by symbols. The solid curves are calculated with Eq.
(\ref{eq:reduced similarity}). The same data are shown on both a
semi-log (a) and a linear (b) scale. The coefficient of determination
is $R^{2}=0.96$.\label{fig:The-similarity-behavior}}

\end{figure}

The similarity behavior in the relaxation process is demonstrated
in Fig. \ref{fig:The-similarity-behavior}. The experimental data
from Figs. \ref{fig:Comparison with RD05}(a) to \ref{fig:Comparison with RD05}(g)
are shifted horizontally and rescaled with $\tau_{2}$. For each case,
$\tau_{2}$ is obtained using $\Gamma_{0}$ listed in table \ref{tab:Listed-parameters}.
The thick solid curve is again the similarity solution from Eq.
(\ref{eq:reduced similarity}), and the results are shown on both
semi-log and linear scales in $\tau$. The coefficient of determination
is $R^{2}=0.96$. The experimental data from a wide range of parameters
demonstrate a universal behavior governed by a single timescale, $\tau_{2}=r_{0}\mu_{e}/\Gamma_{0}$.
This result is a main contribution of the present work. %

We remark that a similar behavior should be observed for droplets,
where the initial membrane tension, $\Gamma_{0}$, is replaced by
$\gamma$, the coefficient of surface tension in $\tau_{2}$ (cf.
the definition of $\tau_{2}$ in Zhang \emph{et al.}).\cite{Zhang2012} However, there
is a subtle difference between droplet and vesicle relaxation while
the coefficient of surface tension is usually a constant, the membrane
tension, $\Gamma_{h}$, is not. Nonetheless, as long as $\Gamma_{h}$
depends linearly on $\Gamma_{0}$, which is a good approximation for
small-to-moderate deformations. The universal behavior in Fig. \ref{fig:The-similarity-behavior}
is expected.%
\begin{figure}
\center

\includegraphics[width=0.5\textwidth]{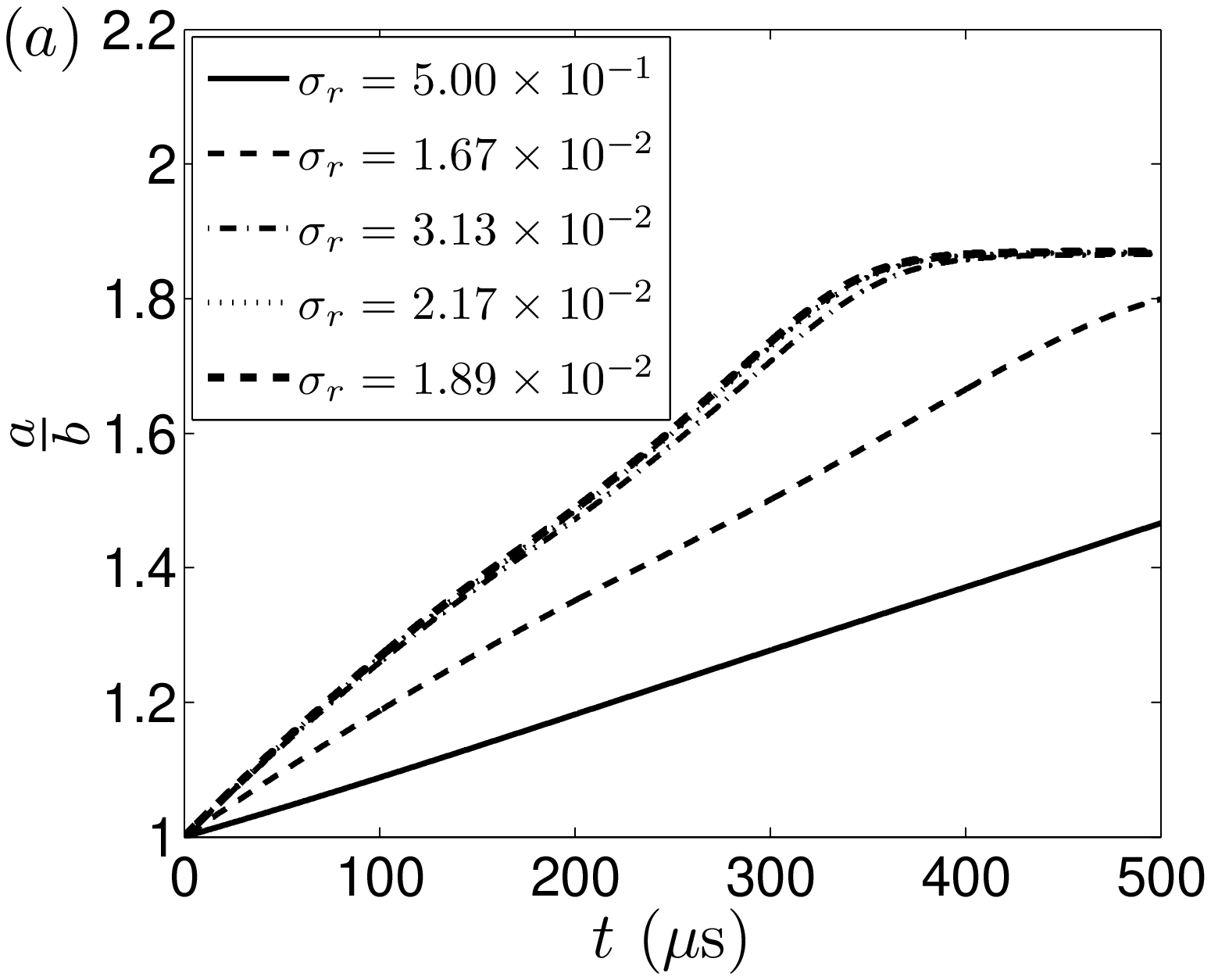}
\includegraphics[width=0.5\textwidth]{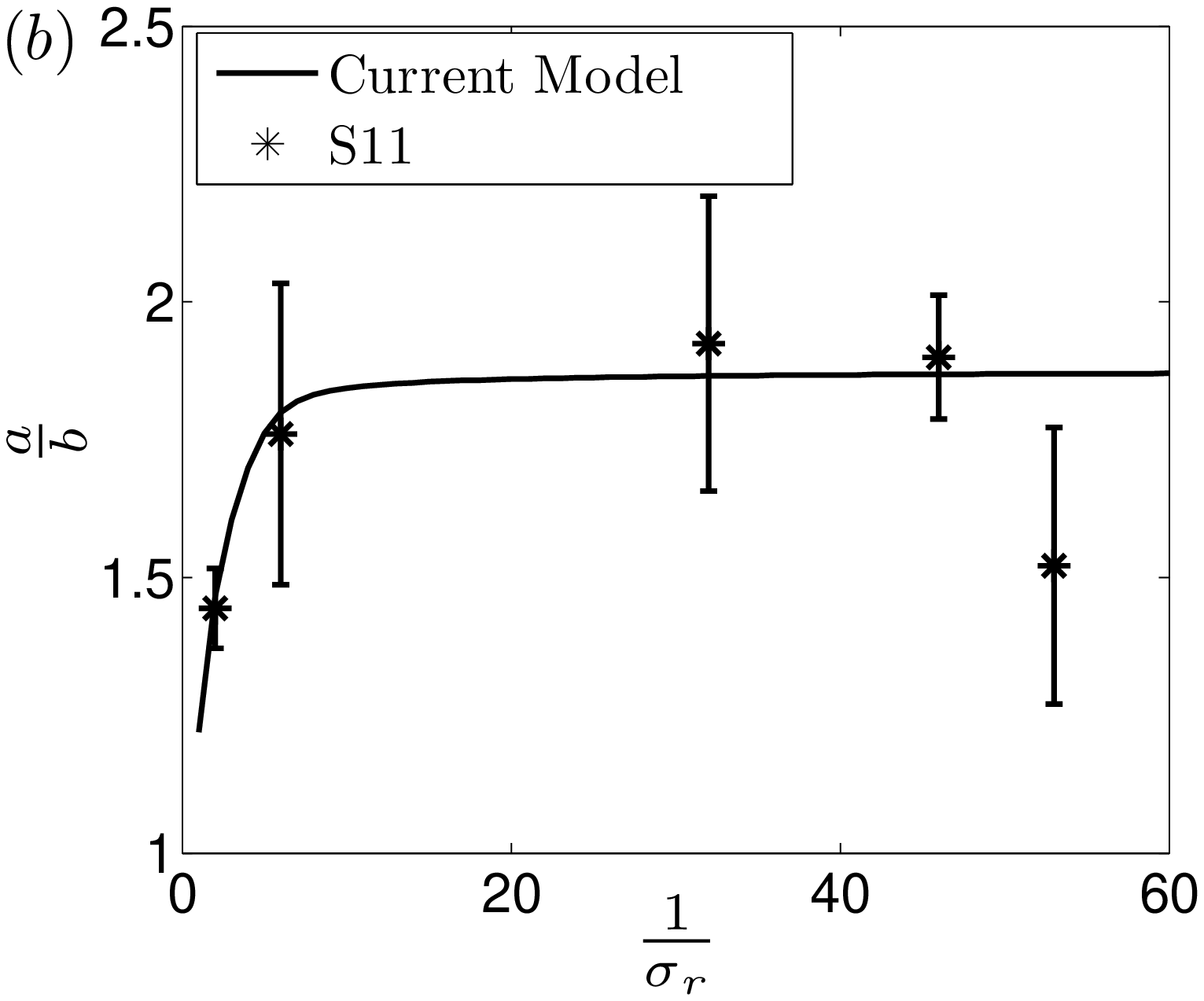}\caption{Comparison with data from S11. (a) Simulated time-course of the aspect
ratio for various conductivity ratios. For all cases $r_{0}=11.3\;{\rm \mu m}$
and $\Gamma_{0}=1\times10^{-8}\;{\rm N/m}$. (b) The aspect ratio
at $t=500\;{\rm \mu s}$ as a function of $1/\sigma_{r}$.\label{fig:Comparison-with-S11}}

\end{figure}

Finally, the model prediction is compared with data from S11. In this
work, the deformation is examined at a fixed pulse length of $t_{p}=500\;{\rm \mu s}$,
and for five intra-to-extra vesicular conductivity ratios. Only the
case of $E_{0}=0.9\;{\rm kV/cm}$ is examined, where no or weak electroporation
is expected. We do not compare the cases of $E_{0}=2$ and 3 kV/cm
in S11, where the vesicles were in the strongly-electroporated regime,
and our model no longer applies. The governing parameters are $r_{0}=11.3\;{\rm \mu m}$
and $\sigma_{e}=3\times10^{-4}\;{\rm S/m}$. The initial membrane
tension is chosen to be the same for all vesicles, namely, $\Gamma_{0}=1\times10^{-8}\;{\rm N/m}$.
Figure \ref{fig:Comparison-with-S11}(a) shows the deformation process
as a function of time for five conductivity ratios. As $\sigma_{r}$
decreases the rate of deformation increases. Except for the case of
$\sigma_{r}=0.5$, the aspect ratio reaches a plateau before the pulse
ends. The time at which the aspect ratio increases saturates with
an increasing $\sigma_{r}$. For $\sigma_{r}=0.5$, an equilibrium
could be reached if the pulse length is extended and sufficiently
long (not shown here). In Fig. \ref{fig:Comparison-with-S11}(b),
the aspect ratio at $t=t_{p}$ is shown as a function of $1/\sigma_{r}$.
We choose this representation to facilitate comparison with the data
from S11 (symbols), where the definition of the conductivity ratio
is $\sigma_{i}/\sigma_{e}$. A reasonable agreement is found between
the two. The behavior of the simulation and the data is explained
by the dependence of the electrical stress on $\sigma_{r}$ in S11
[see Eq. (21) and Sec. 4 therein]. We do not repeat it here for
brevity. The current model represents a significant improvement from
that in S11, where the hydrodynamic problem is treated empirically. 

Some remarks are appropriate before concluding the section. First,
for most cases studied here, the TMP is near the threshold, and the
vesicles are expected to experience no or weak electroporation. For
this regime, our model is shown to provide a good predictive capability,
which demonstrates that the membrane-mechanical model (\ref{membrane tension}),
although derived assuming no electroporation, can be extended to the
weakly-electroporated regime, presumably due to the absence of major
structural alterations. Our model
is not applicable to the strongly-electroporated regime. Second, the
universal scaling law in relaxation observed in Figs. \ref{fig:effect of sigma0},
\ref{fig:effect of tp}, and \ref{fig:The-similarity-behavior} is
expected to hold regardless of the means of deformation, e.g., via
AC/DC electric fields, or via mechanical stretching. Equation (\ref{eq:reduced similarity})
is applicable to a wide range of relaxation phenomena beyond electrodeformation.
Third, the current work suggests that an extensive parametric study
on vesicle electrodeformation-relaxation experimentally, in particular
in the sub-critical regime where electroporation is avoided, can provide
the benefit to further validate our model understanding. A systematic
approach can be possibly developed based on this work to map membrane
properties.

\section{Conclusions}

In this work, we developed a transient analysis for vesicle electrodeformation.
The theory is derived by extending our previous work on a droplet model in Zhang \emph{et al.}, \cite{Zhang2012}
with the additional consideration of a lipid membrane separating two
fluids of arbitrary properties. For the latter, both a membrane-charging
and a membrane-mechanical model are supplied. Similar to the droplet
model, the main result is also an ODE governing the evolution of the
vesicle aspect ratio. The effects of initial membrane tension and
pulse length are examined. The initial membrane tension affects the
relaxation process much more significantly than the deformation process,
in particular when its value is small. The model prediction is extensively
compared with experimental data from Riske and Dimova\cite{Riske2005} and Sadik \emph{et al.},\cite{Sadik2011}
and is shown to accurately capture the system behavior in the regime
of no or weak electroporation. More importantly, the comparison reveals
that vesicle relaxation obeys a universal behavior, and is governed
by a single timescale that is a function of the vesicle initial radius,
the fluid viscosity, and the initial membrane tension. This behavior
is regardless of the means of deformation, either via AC/DC electric
field, or via mechanical stretching. This universal scaling law is
a main contribution of the current work, and can be used to calculate
membrane properties from experimental data. 

\begin{acknowledgements}
JZ and HL acknowledge fund support from an NSF award CBET-0747886 with Dr William Schultz
and Dr Henning Winter as contract monitors.
\end{acknowledgements}

\appendix

\section{}

The functions $f_{14}(\xi_{0})$, $f_{15}(\xi_{0})$,
$f_{21}(\xi_{0})-f_{24}(\xi_{0})$, and $F$ in Eq.
(\ref{vesicle shape evolution}) are given in the following expressions:
\begin{equation}
f_{11}(\xi_{0})=\int\frac{G_{3}(\eta)\eta}{(\xi_{0}^{2}-\eta^{2})}d\eta,\end{equation}
\begin{equation}
f_{12}(\xi_{0})=\frac{1}{\xi_{0}^{2}-1}\left\{ \int\frac{G_{3}(\eta)\eta}{(\xi_{0}^{2}-\eta^{2})}\left(\frac{(1-3\eta^{2})}{(\xi_{0}^{2}-\eta^{2})}-3\right)d\eta\right\} ,\end{equation}
\begin{equation}
f_{13}(\xi_{0})=\frac{G_{3}^{''}(\xi_{0})G_{5}^{'}(\xi_{0})-G_{3}^{'}(\xi_{0})G_{5}^{''}(\xi_{0})}{2N}\cdot f_{11}(\xi_{0}),\end{equation}
\begin{equation}
f_{14}(\xi_{0})=-\xi_{0}H_{3}^{'}(\xi_{0})\int\frac{G_{3}(\eta)\eta}{(\xi_{0}^{2}-\eta^{2})^{2}}d\eta+\frac{1}{2}H_{3}^{''}(\xi_{0})f_{11}(\xi_{0}),\end{equation}
\begin{widetext}
\begin{equation}
f_{15}(\xi_{0})=-\frac{H_{3}^{'}(\xi_{0})\left[G_{3}(\xi_{0})G_{5}^{''}(\xi_{0})-G_{3}^{''}(\xi_{0})G_{5}(\xi_{0})\right]}{2N}f_{11}(\xi_{0})+\xi_{0}H_{3}^{'}(\xi_{0})\int\frac{G_{3}(\eta)\eta}{(\xi_{0}^{2}-\eta^{2})^{2}}d\eta.\end{equation}
Here $N\equiv G_{3}(\xi_{0})G_{5}^{'}(\xi_{0})-G_{3}^{'}(\xi_{0})G_{5}(\xi_{0})$. $G$ and $H$ are Gegenbauer functions of the first and
second kind, respectively. The detailed expressions for $G$ and $H$ are found in Dassios \emph{et al.}.\cite{Dassios1994}

\begin{equation}
f_{21}(\xi_{0})=\frac{1}{2}\xi_{0}^{2}\int\frac{(\eta^{2}-1)(3\eta^{2}-1)}{(\xi_{0}^{2}-\eta^{2})}d\eta,\end{equation}
\begin{equation}
f_{22}(\xi_{0})=\xi_{0}f_{11}(\xi_{0})\left[-H_{3}^{'}(\xi_{0})\int\frac{(1-3\eta^{2})(\xi_{0}^{2}-3\xi_{0}^{2}\eta^{2}+2\eta^{4})}{(\xi_{0}^{2}-\eta^{2})^{2}}d\eta+3\xi_{0}H_{3}(\xi_{0})\int\frac{1-3\eta^{2}}{(\xi_{0}^{2}-\eta^{2})}d\eta\right],\end{equation}
\begin{equation}
f_{23}(\xi_{0})=\xi_{0}f_{11}(\xi_{0})\left[-\frac{49(1-3\xi_{0}^{2})G_{3}(\xi_{0})H_{3}^{'}(\xi_{0})}{30N}+H_{3}^{'}(\xi_{0})\int\frac{(1-3\eta^{2})(\xi_{0}^{2}-3\xi_{0}^{2}\eta^{2}+2\eta^{4})}{(\xi_{0}^{2}-\eta^{2})^{2}}d\eta\right],\end{equation}
\begin{equation}
f_{24}(\xi_{0})=\xi_{0}^{3}(1-\xi_{0}^{-2})^{\frac{5}{6}}\int\frac{3\eta^{2}-1}{(\xi_{0}^{2}-\eta^{2})^{\frac{3}{2}}}d\eta+\xi_{0}(1-\xi_{0}^{-2})^{-\frac{1}{6}}\int\frac{3\eta^{2}-1}{\sqrt{\xi_{0}^{2}-\eta^{2}}}d\eta,\end{equation}
\begin{equation}
F=-\frac{2}{3}\left(f_{25}(\xi_{0})+f_{26}(\xi_{0})/\mu_{r}\right),\end{equation}
where
\begin{equation}
f_{25}(\xi_{0})=-\frac{f_{22}(\xi_{0})}{\xi_{0}f_{11}(\xi_{0})}\frac{(\mu_{r}-1)f_{12}(\xi_{0})+f_{13}(\xi_{0})}{\mu_{r}f_{14}(\xi_{0})+f_{15}(\xi_{0})}-3\xi_{0}\int\frac{3\eta^{2}-1}{(\xi_{0}^{2}-\eta^{2})}d\eta-\frac{\xi_{0}}{\xi_{0}^{2}-1}\int\frac{(2\xi_{0}^{2}-\eta^{2}-1)(1-3\eta^{2})^{2}}{(\xi_{0}^{2}-\eta^{2})^{2}}d\eta,\end{equation}
\begin{equation}
f_{26}(\xi_{0})=-\frac{f_{23}(\xi_{0})}{\xi_{0}f_{11}(\xi_{0})}\frac{(\mu_{r}-1)f_{12}(\xi_{0})+f_{13}(\xi_{0})}{\mu_{r}f_{14}(\xi_{0})+f_{15}(\xi_{0})}-\frac{49(1-3\xi_{0}^{2})G_{3}^{'}(\xi_{0})}{30N}+\frac{\xi_{0}}{\xi_{0}^{2}-1}\int\frac{(2\xi_{0}^{2}-\eta^{2}-1)(1-3\eta^{2})^{2}}{(\xi_{0}^{2}-\eta^{2})^{2}}d\eta.\end{equation} $\mu_{r}\equiv\mu_{e}/\mu_{i}$ is the viscosity ratio.
\end{widetext}

\bibliography{reference_on_Vesicle_deformation}

\end{document}